\documentclass[sigconf]{acmart}

\newcommand{\sys}{Antler\xspace}

\usepackage[linesnumbered,ruled]{algorithm2e}
\SetKw{KwAnd}{and}
\SetKw{KwAlgo}{Algorithm}

\usepackage{xspace}
\usepackage{subfiles}
\usepackage{mathtools}
\usepackage{comment}
\usepackage{subfig}
\usepackage{booktabs}
\usepackage{wrapfig}
\usepackage{multirow}
\usepackage{dashrule}
\usepackage{hyperref}
\usepackage{todonotes}
\usepackage{soul}

\setcopyright{none}
\renewcommand\footnotetextcopyrightpermission[1]{} % removes footnote with conference information in first column
\usepackage{pifont}
\newcommand{\parlabel}[1]{\vspace{0.5em}\noindent\textbf{#1}}

\usepackage{amsmath}

\AtBeginDocument{%
  \providecommand\BibTeX{{%
    \normalfont B\kern-0.5em{\scshape i\kern-0.25em b}\kern-0.8em\TeX}}}

\acmYear{2022}
\acmConference[Conference '23]{The ACM/IEEE International Conference on Information Processing in Sensor Networks}
\acmISBN{978-1-4503-XXXX-X/18/06}

\begin{document}


%\title{SmartSwitch: Enable Efficient Context Switch for DNNs on Embedded Systems}

\title{Efficient Multitask Learning on Resource-Constrained Systems}

\author{Yubo Luo, Le Zhang, Zhenyu Wang, Shahriar Nirjon}
% \email{yubo@cs.unc.edu}
\affiliation{%
  \institution{University of North Carolina}
  % \streetaddress{P.O. Box 1212}
  \city{Chapel Hill}
  \state{NC}
  \country{USA}
}

% \affiliation{%


\begin{abstract}
We present \sys, which exploits the affinity between all pairs of tasks in a multitask inference system to construct a compact graph representation of the task set and finds an optimal order of execution of the tasks such that the end-to-end time and energy cost of inference is reduced while the accuracy remains similar to the state-of-the-art. The design of \sys is based on two observations: first, tasks running on the same platform shows affinity, which is leveraged to find a compact graph representation of the tasks that helps avoid unnecessary computations of overlapping subtasks in the task set; and second, tasks that run on the same system may have dependencies, which is leveraged to find an optimal ordering of the tasks that helps avoid unnecessary computations of the dependent tasks or the remaining portion of a task. We implement two systems: a 16-bit TI MSP430FR5994-based custom-designed ultra-low-power system, and a 32-bit ARM Cortex M4/M7-based off-the-shelf STM32H747 board. We conduct both dataset-driven experiments as well as real-world deployments with these systems.  We observe that \sys's execution time and energy consumption are the lowest compared to all baseline systems and by leveraging the similarity of tasks and by reusing the intermediate results from previous task, \sys reduces the inference time by 2.3X -- 4.6X and saves 56\% -- 78\% energy, when compared to the state-of-the-art.

\end{abstract}




\settopmatter{printfolios=true}
\maketitle

\section{Introduction}

%~\cite{spyridon2012classification,prassler2008domestic,gates2007robot, bohren2011towards,siegwart2003robox}

In recent years, we see an increased number of low-resource systems that are running deep neural networks under extreme CPU, memory, time, and energy constraints~\cite{zygarde,yao2017deepsense,yao2018deep,yao2017deepiot,lee2019intermittent}. Nowadays, it is becoming common to see multiple neural networks co-residing on the same portable, wearable, and mobile edge device in order to offer a wide variety of intelligent applications and services to the user~\cite{kawsar2018esense,islam2019device}. Many IoT devices have built-in voice assistants that authenticate the speaker, understand what they say, and recognize gestures, facial expressions, and emotions. Mobile vision technology~\cite{redmon2015real,xie2017towards,liao2016understand,sarikaya2017detection,giusti2015machine,smarton,spoton} is built into many mobile and social robots that perform on-device object recognition, obstacle detection, scene classification, localization, and navigation. In order to increase the accuracy and robustness of these classifiers, numerous \emph{multitask learning} techniques have been proposed in the mainstream machine learning literature~\cite{ruder2017overview,sun2020adashare}. Some of these techniques have been adopted by the embedded systems community to scale up the number of classifiers that co-exist on an embedded system~\cite{nwv, yono}.

Unfortunately, multitask learning on low-resource embedded systems still remains a challenge. Slow CPU, scarce memory (RAM), and high overhead of external storage (flash) make the response time and the energy cost of multitask inference on these systems extremely high. To deal with these challenges, recent works~\cite{nwv, yono, nws} have proposed bold measures such as squeezing all~\cite{nwv, yono} or most~\cite{nws} of the neural networks into the main memory (RAM) --- in order to avoid the high overhead of storage and to rely on fast, in-memory computation for the most part. However, speedup gained in such extreme ways inevitably comes at the cost of lower accuracy and/or hidden time and energy cost that overshadows the benefit of in-memory execution. In general, state-of-the-art multitask inference techniques for low-resource systems lack two major aspects that could significantly reduce the inference time and energy consumption: 

Firstly, inference tasks that run on the same system generally show affinity. For instance, a speaker identification task and a speech recognition task running on a voice assistant device share common latent subtasks such as noise compensation and phoneme identification. These overlapping subtasks should be factored out and executed only once to reduce the time and energy waste due to repeating them for both tasks. Existing works~\cite{nwv, yono, nws} pack multiple tasks in the main memory by sharing task constructs at the granularity of weights. They do not exploit the higher-level affinity between tasks and thus fail to recognize that even though in-memory operations are faster, repeatedly executing subtasks adds up significantly higher overhead, especially when it involves multiple convolutional layers, which is completely unnecessary.

%At one hand, these tasks consume information from the same sensor, and on the other hand, they deliver results to the same user or application. 

Secondly, inference tasks that run on the same system generally manifest inter-task and intra-task dependencies -- requiring the system to execute tasks and subtasks in a certain order. This also creates opportunities to skip a dependent task or a subtask. For instance, voice classification tasks are routinely preceded by a lightweight voice activity detector to reduce computational overhead. Likewise, subtasks such as noise compensation and phoneme detection often precede the rest of the audio processing pipeline in typical speech classification tasks. Existing works~\cite{nwv, yono, nws} that merge or load tasks based on the byte-values of the weights are not capable of exploiting higher-level inter-task and intra-task dependencies, and thus they waste time and energy in executing tasks and subtasks that are unnecessary.

In this paper, we introduce \sys -- which exploits the affinity between tasks in a multitask inference system to construct a compact graph representation of the task set. Unlike existing task grouping techniques that are primarily concerned with inference accuracy only, we construct task graphs considering both the accuracy as well as the time and/or energy waste from repeated execution of subtasks. Furthermore, we observe that different pairs of tasks exhibit different degrees of affinity and the cost of switching from one task to another is nonidentical. We formally prove that ordering tasks in a multitask learning scenario is NP-Complete and provide an integer linear programming formulation to solve it. We extend the formulation to include dependency constraints between tasks and subtasks. We describe a genetic algorithm to solve the optimization problem for both constrained and unconstrained cases.

\begin{figure*}[!t]
    \centering
    \includegraphics[width=6.5in]{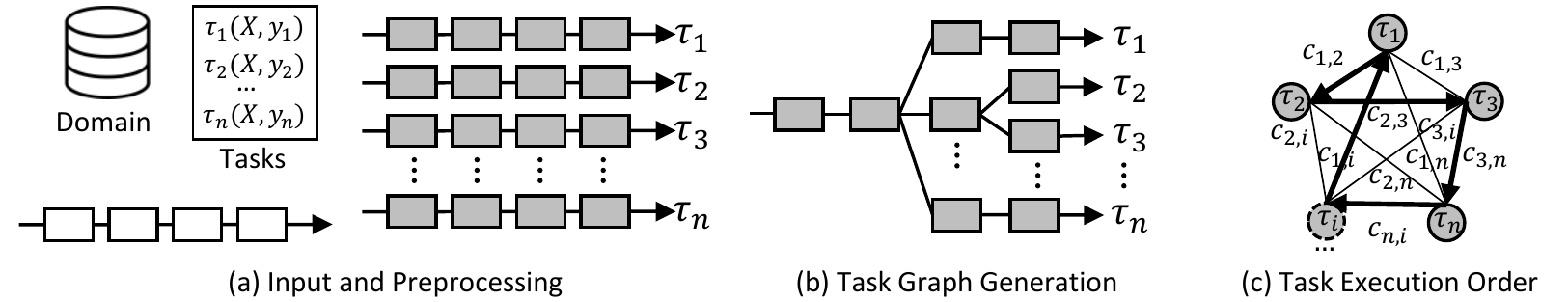}
    \caption{Overview of \sys: (a) A set of tasks defined over a domain is taken as the input. A common network architecture is individually trained to produce network instances (one for each task). (b) A task graph is formed considering both accuracy and task execution cost. The task graph is retrained. (c) An optimal task execution order minimizes the task execution cost.}
    \label{fig:sys_overview}
\end{figure*}

In order to evaluate \sys, we develop two systems: 1) a 16-bit TI MSP430FR5994-based custom-designed ultra-low-power system, and 2) a 32-bit ARM Cortex M4/M7-based off-the-shelf STM32H747 board. We conduct dataset-driven experiments as well as real-world deployments with these systems. In the dataset-driven experiments, we compare the performance of \sys against four baseline solutions, including three state-of-the-art multitask inference systems for low-resource systems: YONO~\cite{yono}, NWV~\cite{nwv} and NWS~\cite{nws} over nine datasets that are used in the literature. We observe that \sys's execution time and energy consumption is the lowest compared to all baseline systems. By leveraging the similarity of tasks and by reusing the intermediate results from previous task, \sys reduces the inference time by 2.3X -- 4.6X and saves 56\% -- 78\% energy, when compared to the baselines. In the real-world deployments, we implement two multitask inference systems having five audio and four image inference tasks. Results show that \sys reduces the time and energy cost by 2.7X -- 3.1X while its inference accuracy is similar to running individually-trained classifiers within an average deviation of $\pm 1\%$.

\section{Overview of \sys}

\sys is a tool for developing efficient multitask deep learning models for low-resource systems that have extreme CPU, memory, and energy constraints. This section provides an overview of \sys, deferring its technical details to later sections.      

\subsection{Input and Preprocessing}

\parlabel{Tasks.} \sys takes a set of inference tasks defined over a domain as the input. Formally, for a given set of tasks, $\mathrm{\tau = \{\tau_1(X, y_1), \tau_1(X, y_2),}$ $\mathrm{\cdots,  \tau_1(X, y_n)\}}$ defined over the domain $\mathrm{X}$, each task $\mathrm{\tau_i}$ maps a sample, $\mathrm{x_j \in X}$ to a class label, $\mathrm{y_{(j,i)}}$, where $\mathrm{y_i = [y_{(1, i)}, \cdots y_{(n, i)}]}$.       

For example, over an audio dataset, $\mathrm{X}$, we can define three tasks:  a speaker recognition task ($\mathrm{\tau_1}$),  a speech recognition task ($\mathrm{\tau_2}$), and an emotion classification task ($\mathrm{\tau_3}$). For each audio clip, $\mathrm{x \in X}$ these three classifiers output the class labels for the speaker, the speech, and the emotion, respectively.  

%A neural network inference \emph{task} takes sensor data samples as the input and classifies it as one of a fixed set of predefined categories as the output. In a multitask learning scenario, we assume a predefined set of tasks, $\mathrm{\tau = \{\tau_1(X, y_1), \tau_1(X, y_2), \cdots \tau_1(X, y_n)\}}$ defined over a domain $\mathrm{X}$. A classifier, $\mathrm{\tau_i}$ maps a sample, $\mathrm{x_j \in X}$ to a class label, $\mathrm{y_{(j,i)}}$, where $\mathrm{y_i = [y_{(1, i)}, \cdots y_{(n, i)}]}$.   

% \begin{figure}[!htb]
%     \centering
%     \includegraphics[scale=0.9]{figure/ppt/tasks.pdf}
%     \caption{Tasks and neural networks.}
%     \label{fig:tasksnetworks}
%     %\vspace{-20pt}
% \end{figure}

\parlabel{Preprocessing.} For each task, \sys instantiates a neural network. \sys uses a common network architecture for all tasks, which is trained on the dataset, $\mathrm{(X, y_i)}$ to instantiate the neural network corresponding to each task $\mathrm{\tau_i}$. A common network architecture is necessary for \sys since its ultimate goal is to form a multitask neural network that consists of two or more networks sharing one or more of their network layers. A common network architecture is methodologically obtained by running a network architecture search~\cite{elsken2019neural} that empirically optimizes the accuracy of all networks. To speed up the search, we start from a library of popular neural network architectures from the literature ~\cite{nws, nwv, yono} and run a hyper-parameter search to obtain the common network architecture that maximizes the minimum accuracy of all tasks. Figure~\ref{fig:sys_overview}(a) shows the network instances (one for each task) which have identical architecture but different weights.     

%\vspace{-10pt}
\subsection{Task Graph Generation}
%\subsection{Task Graph}
%\label{sec:taskgraph}

\parlabel{Task Graph.} Tasks in a multitask learning scenario share parts of their networks to influence each other during their joint training. Typically, tasks share their first few layers since layers closer to the input tend to encode simpler basic patterns which are the building blocks for similar inference tasks. For example, early layers of an audio classifier for human voice encodes phonemes and morphemes that are building blocks to downstream tasks such as keyword spotting, speech recognition, and sentiment analysis.

Different pairs of tasks may share different number of layers depending on how similar the tasks are. In \sys, this is represented by a tree-like structure, which we call a \emph{task graph}, as shown in Figure~\ref{fig:sys_overview}(b). Each rectangular box in the figure represents a \emph{block} which consists of one or more layers. A path from the root (i.e., the leftmost block) to a leaf (i.e., one of the rightmost blocks) corresponds to one neural network inference task. Notice that a block may be shared by two or more neural networks. 

%Too compact:

\parlabel{Task Graph Generation.} \sys analyzes the \emph{affinity} between the network instances to form a task graph that balances the trade-off between having a task graph that is too compact vs. having a task graph that has little to no overlap between tasks. 

Compact task graphs are generally desirable since they require less storage, save time and energy by avoiding repeated computations, and take advantage of multitask learning such as reduced overfitting and knowledge transfer which is facilitated by the shared network structures. Compact task graphs, however, generally have less network capacity due to less number of parameters, which limits their ability to accurately classify large and complex data. Section~\ref{sec:taskgraphgeneration} describes how \sys finds an optimal task graph that balances these two opposing forces. 

Once a task graph architecture is decided, all tasks are re-trained using a standard multitask learning algorithm~\cite{vandenhende2019branched}.  

%increase time and/or energy waste, and too sparse task graphs that not only require more storage space but also fail to take the full advantage of multitask learning which reduces over-fitting through the use of shared representation. 

% \begin{figure}[!htb]
%     \centering
%     \includegraphics[scale=.9]{figure/ppt/taskgraph.pdf}
%     \caption{Task graph.}
%     \label{fig:taskgraph}
%     %\vspace{-20pt}
% \end{figure}

\subsection{Task Execution Order}

\parlabel{Task Graph Execution Process.} In memory-constrained systems, neural networks are executed progressively in multiple stages. Depending on the size of the main memory, network weights and parameters corresponding to one or more layers are brought into the main memory from the non-volatile storage (e.g., flash) prior to the execution of those layers. Hence, the total cost of executing the tasks in a task graph depends not only on the number of blocks the task graph contains but also on how often each block is brought into the main memory for execution.   

In \sys, all tasks have the same network architecture. When the system starts up, memory is statically allocated in the RAM having the size of the common network architecture. At runtime, prior to the execution of each task $\tau_i$, blocks containing its weights and parameters are loaded into the main memory to initialize the common network architecture -- which is then executed to obtain the inference result. Because tasks in \sys can share blocks, \sys skips loading the blocks that are already in the main memory in order to reduce the read/write overhead. Additionally, intermediate results after executing each block are cached in memory buffers (one buffer after each block) to avoid repeated computation of blocks.  

\parlabel{Optimal Task Execution Order.} Since in-memory blocks are not reloaded or re-executed if the next task needs them and since different pairs of tasks generally share different number of blocks, the order in which tasks are executed affects the total cost of executing the tasks. In Figure~\ref{fig:sys_overview}(c), the overhead of switching from one task to another is represented by a weighted complete graph whose nodes represent tasks and weights, $c_{i,j}$ on each edge represent the cost of switching between tasks. Finding the least cost ordering of the tasks therefore is equivalent to finding a least-cost Hamiltonian cycle (shown with arrows) on this graph~\cite{papadimitriou1998combinatorial}. 

Furthermore, tasks may have precedence constraints and conditional dependencies between them. These add additional constraints on their execution order. Section~\ref{sec:taskorder} describes how \sys finds an optimal ordering of tasks for a given task graph where tasks may have ordering constraints.

\section{Task Graph Generation}
\label{sec:taskgraphgeneration}

This section describes how task graphs are generated from the network instances obtained after the preprocessing step.   
                  
\subsection{Quality of a Task Graph}
\label{sec:affinitytaskgraph}

%\parlabel{Task Affinity.} Task affinity is a measure of how similar two tasks are in a multitask learning scenario. This helps us determine which tasks can be grouped together and are likely to share one or more network layers. A high affinity score between two tasks mean they can share several early layers and thus result in a compact task graph. A low affinity score means the otherwise. 
%An illustration of two extreme cases of task affinity are shown in Figure~\ref{fig:taskaffinitydifferent}.

%Several methods have been proposed in the literature to quantify the similarity between two machine learning tasks ~\cite{dwivedi2019,zamir2018taskonomy,lu2017fully,standley2020tasks,vandenhende2019branched}. Instead of proposing a new method, we adopt ~\cite{vandenhende2019branched,dwivedi2019}, which implements a globally optimal grouping strategy unlike other methods that employ greedy algorithms to find a locally optimal grouping. We should clarify that although we borrow the task affinity metric from ~\cite{vandenhende2019branched,dwivedi2019}, our grouping algorithm is quite different from existing literature since we are the first to consider both task affinity and execution cost in multitask learning.

%For each network, we can obtain a representation of the input data after executing each layer.

\parlabel{Task Affinity.} Task affinity refers to the degree at which two tasks are similar in their data representation~\cite{vandenhende2019branched,dwivedi2019}. For a pair of tasks in \sys, we should ideally compare the outputs of every layer over the entire dataset. This is, however, costly and is also not necessary in practice. Hence, to lower the computational overhead, we choose $\mathrm{D}$ layes, which are referred to as \emph{branch points}, and measure the similarity of outputs of the two networks at these branch points over a subset of $\mathrm{K}$ random samples from the dataset. Computing task affinity is a two-step process: 

\textbf{Step 1 --} Each task is profiled using $\mathrm{K}$ data samples. At each branch point, for all pairs of samples, the dissimilarity of their representations is computed using inverse Pearson correlation coefficient~\cite{taylor1990interpretation} to obtain a $\mathrm{D \times K \times K}$ dimensional tensor. This tensor is flattened to have a vector that encodes the data representation profile of a task. This process is repeated for each task.  

\textbf{Step 2 --}  Affinity score for each pair of tasks is computed. At each branch point, for all pairs of tasks, the similarity of their data representation profile tensors is computed using Spearman's correlation coefficient~\cite{taylor1990interpretation} to obtain a $\mathrm{D \times n \times n}$ dimensional matrix where $\mathrm{n}$ is the number of tasks. This matrix encodes the similarity between pair of tasks at each branch point. This information is used later when tasks are grouped to form affinity-aware task graphs.

\parlabel{``Variety'' Score of Task Graphs.} We extend the definition of task affinity to task graphs. The subtree rooted at each branch point of a task graph contains a subset of tasks that share one or more blocks. In other words, all the blocks from the root of the graph to the root of the subtree are shared by all tasks that are on the leaf of the subtree. At the root of the subtree, tasks diverge and follow different paths. We quantify this using \emph{variety} score.   

We define \emph{variety} score for the tasks under each branch point as the average maximum dissimilarity score over all pair of tasks under that branch point. The variety score quantifies the dissimilarity or misfit within tasks under each branch point. The overall variety score of a task graph is the sum of variety scores at all branch points. Computing the variety score of a task graph is a two-step process: 

\textbf{Step 1 --} Variety score at each branch point is computed using Equation~\ref{eq:sl}, where $\mathrm{S_{\rho, i,j}}$ denotes the affinity between tasks $\mathrm{\tau_i}$ and $\mathrm{\tau_j}$ at branch point $\mathrm{\rho}$, $\mathrm{c_k}$ denotes the $\mathrm{k}$-th child branch, and $\mathrm{m}$ denotes the total number of child branches.     
\begin{align}
\mathrm{v_{\rho} = \frac{1}{m} \left[\sum_{k=1}^m \max_{ i,j \in c_k}\big(1-S_{\rho,i,j}\big)\right]} \label{eq:sl}
\end{align}

\textbf{Step 2 --} Variety score, $\mathrm{v_{\rho}}$ from all branch points are summed to obtain the variety score for the task graph using Equation~\ref{eq:ft}.    
\begin{align}
\mathrm{V = \sum_{\rho} v_{\rho}} \label{eq:ft}
\end{align}

Note that although we use \emph{affinity}, a similarity score to quantify the similarity between two tasks, we use \emph{variety} score, a measure of dissimilarity to quantify a task graph's quality. This may seem counter intuitive but this is similar to intra-cluster distance in clustering algorithms which measures the impurity in a cluster. 

\begin{figure}[!htb]
    \centering
    \includegraphics[scale=.9]{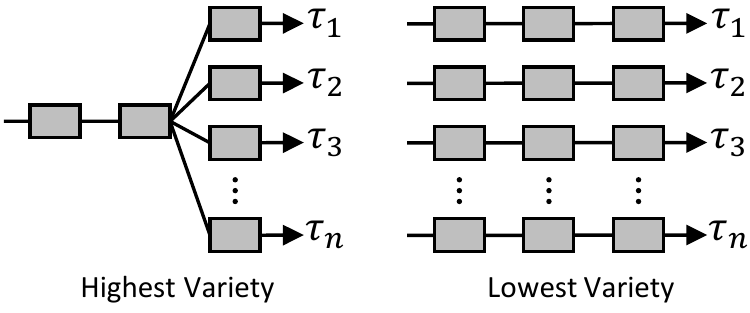}
    \caption{Examples of task graphs having very high (left) and very low (right) variety scores.}
    \label{fig:taskaffinitydifferent}
    %\vspace{-20pt}
\end{figure}

\subsection{Tradeoff Analysis}
\label{sec:tradeoff}

Task graphs with low variety scores are generally desired as variety score tends to inversely correlate with inference accuracy. However, the lower the variety score of a task graph is, the higher its time, energy, and storage overhead is.

For example, the task graph in Figure~\ref{fig:taskaffinitydifferent} {(left)} has the highest variety score --- all tasks are essentially in one group. This is the most compact representation for any task set and has several benefits such as the least storage requirement and the least time and energy overhead when switching tasks. However, since all tasks share pretty much all layers, the likelihood of individual task performing accurate inferences is low. 

On the other hand, the task graph in Figure~\ref{fig:taskaffinitydifferent} {(right)} has the lowest variety score --- each task forms its own group, has the maximum time, energy, and storage overhead (as no blocks are shared), but since each task retains its weights, the inference accuracy is likely to be relatively higher.        

%\parlabel{Variety Score vs. Execution Cost Tradeoff.} 

%Low-resource embedded systems lack capacity to execute an arbitrarily large sized neural network. CPU and memory constraints set a limit on the maximum sized multitask inference graph that these systems can execute. Hence, our goal is to find an optimal task graph that maximizes the inference accuracy, minimizes the task execution cost, while its size does not exceed a given budget.                   

\begin{figure}[!htb]
    \includegraphics[width=0.3\textwidth]{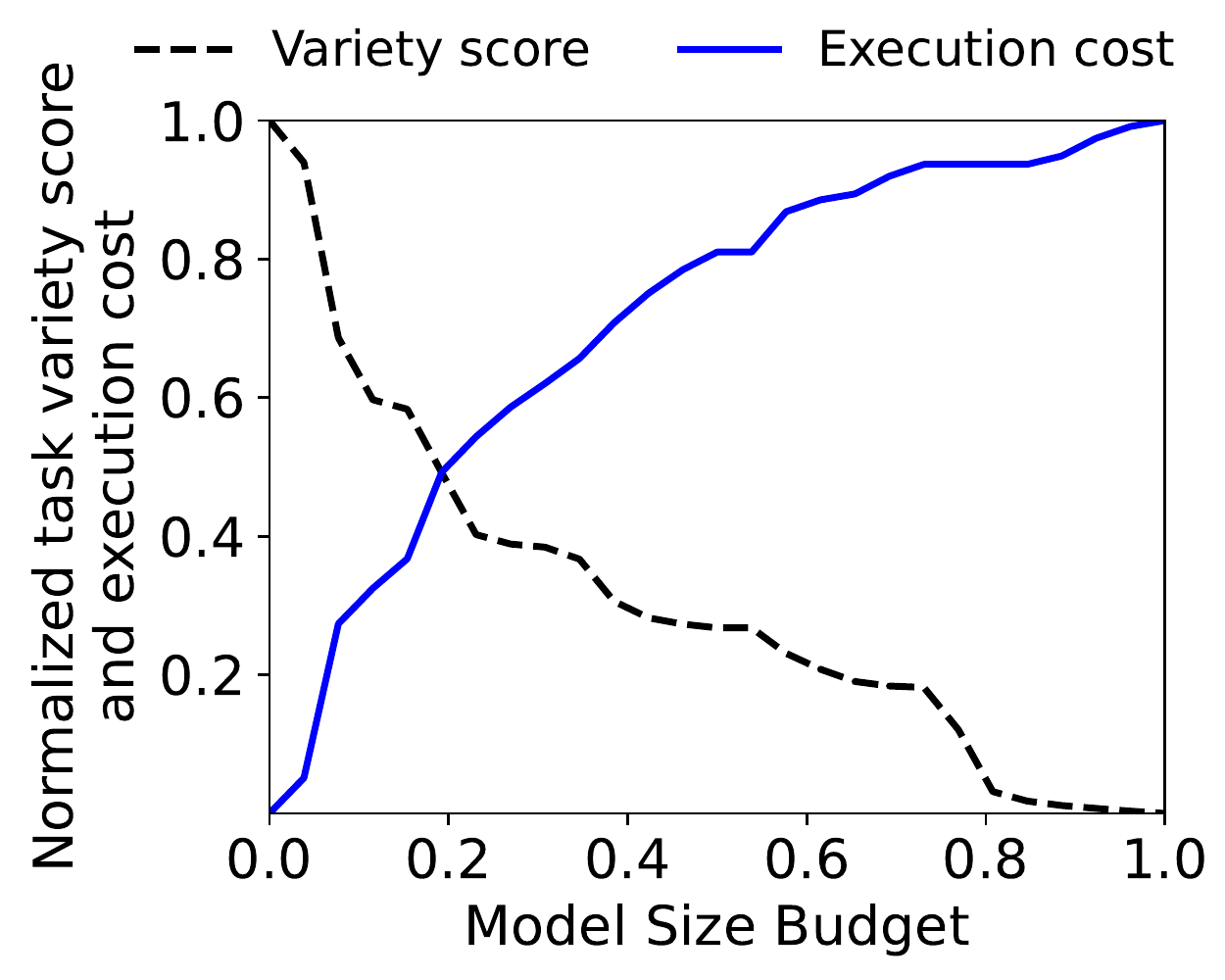}
    \caption{Tradeoff between variety score and execution cost.}
    %All values are normalized by max-min normalization method. 
    \label{fig:algo1_tradeoff}
\end{figure} 

%The two goals --- high accuracy and low execution cost --- however, are not achievable at the same time. 

\parlabel{Empirical Tradeoff Curve.}  Figure~\ref{fig:algo1_tradeoff} shows this tradeoff using empirical data obtained from one of our experiments. We define five image classification tasks on the dataset~\cite{lecun1998gradient} and use a five-layer CNN having 2 convolutional and 3 fully-connected layers as the common network architecture. We generate all possible task graphs, compute their variety scores, estimate their execution costs, and record their model sizes. 

To draw the tradeoff curve, we vary the maximum model size budget, and for each budget, we pick the task graph having the lowest variety score and whose size is within the budget. The variety score and the execution cost of that task graph are normalized and plotted on the Y-axis. Thus, we get trend lines for variety score and execution cost.         

We observe that although an increased model size budget allows us to have a task graph with lower variety score, it comes at the cost of increased execution overhead. To balance these two opposing goals, \sys chooses the task graph that lies at the intersection of the two trend lines.   

%In Figure~\ref{fig:algo1_tradeoff},  where the variety score and the execution cost are shown for different model size budgets.

%For each model size budget, we generate all possible task graphs, and compute their variety scores as well as task execution costs. For each budget, we plot the minimum variety score and the minimum execution cost to obtain the two trend lines. We observe that as the budget constraint is relaxed, the variety score decreases (i.e., expected inference accuracy increases) while the execution cost increases.   

%In the experiment above, we use a 5-layer neural network having 2 convolutional and 3 fully-connected layers as the common network architecture and define 5 image classification tasks on the dataset~\cite{lecun1998gradient}. 

%For different multitask inference problems, the exact values on this plot are different, but the general trend remains the same. We call this the \emph{variety-cost trade-off curve} which is used to select task graphs that are accurate and run under the size and overhead budgets. 

\subsection{Task Graph Generation Algorithm}

Given $\mathrm{n}$ individually trained neural networks having the same architecture, generating the task graph is a five-step process:

\textbf{Step 1 --} For each pair of tasks, their affinity score is computed at $\mathrm{D}$ branch points to obtain a $\mathrm{D \times n \times n}$ matrix. 

\textbf{Step 2 --} The set of all task graphs containing $\mathrm{n}$ tasks, $\mathrm{G_T(n)}$ is generated through a recursive process. For every task graph $\mathrm{g \in G_T(n-1)}$, where $\mathrm{g}$ contains $\mathrm{n-1}$ tasks, we generate $\mathrm{\Lambda(g)}$ new task graphs, each containing $\mathrm{n}$ tasks. $\mathrm{\Lambda(g)}$ denotes the number of non-leaf internal node of $\mathrm{g}$. This is because the $\mathrm{n}$-th task can only branch out from one of the non-leaf internal nodes of $\mathrm{g}$.

\textbf{Step 3 --} For each task graph, $\mathrm{g \in G_T(n)}$, its variety score, model size, and execution cost are estimated. The variety score is obtained using Equation~\ref{eq:ft}. The execution cost is estimated from empirical measurements of the cost of executing each block of the common network architecture. Execution cost estimation also requires the optimal execution order of the tasks, which is obtained using the algorithm described in the next section.

\textbf{Step 4 --} The variety score vs. execution cost tradeoff curve is computed. The task graph, $\mathrm{\hat{g}}$ corresponding to the point where the two trend lines intersect is returned.       

\textbf{Step 5 --} The selected task graph, $\mathrm{\hat{g}}$ is retrained using ~\cite{vandenhende2019branched}.     

% \begin{algorithm}[!t]

% \small
%  N tasks are individually trained;\\
%  D branch points are selected; \\
%  \For{each task}{
%     K images are given to get the output of the D branch points; \\ 
%     a $\mathrm{D \times K \times K}$ tensor is obtained;\\
%  }
%  \For{each task pair}{
%     affinity score is computed at each branch point;\\
%  }
%  a $\mathrm{D \times N \times N}$ matrix is obtained; \\
%  \For{each model size budget}{
%     find the point with minimum variety score;\\
%     find the point with minimum execution cost;\\
%  }
%  select the trade-off model size budget; \\
%  determine the graph that yields the minimum execution cost;
%  \caption{Task Graph Generation}\label{algo:algorithm1}
% \end{algorithm}

%\newpage
\section{Optimal Task Execution Order}
\label{sec:taskorder}

This section describes how \sys achieves the optimal task ordering for a given task graph that may or may not have precedence or conditional constraints. This problem is NP-complete and a proof is in Appendix~\ref{sec:appendix1}.         

\subsection{Task Execution Order}

\parlabel{Significance.} Task graphs provide a compact representation of tasks showing the overlaps in them. These graphs do not explicitly impose any order for executing the tasks. We observe that for a given task graph with $\mathrm{n}$ tasks, not all $\mathrm{n!}$ permutations of their execution order are the same. This is because different pairs of tasks in a task graph generally share different number of blocks. 

\begin{figure}[!htb]
    \centering
    \includegraphics[scale=1]{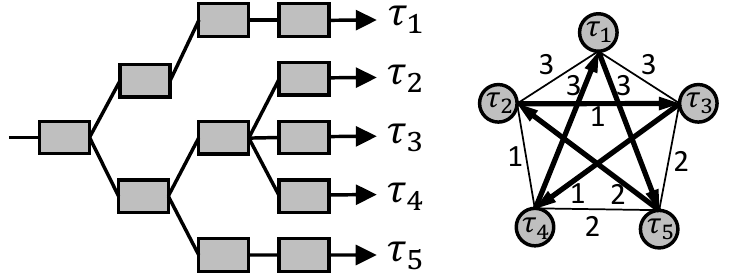}
    \caption{Switching cost is different for different task pairs.}
    \label{fig:taskorderingexample}
    %\vspace{-20pt}
\end{figure}

Consider the task graph with five tasks in Figure~\ref{fig:taskorderingexample} as an example. The cost of switching from one task to another is shown on the weighted complete graph on its right. For simplicity of illustration, we assume the cost of loading and executing each block is 1 unit. We observe that executing the tasks in the order: $\mathrm{\tau_2 \rightarrow \tau_1 \rightarrow \tau_3 \rightarrow \tau_5 \rightarrow \tau_4}$ incurs significantly higher overhead when compared to the optimal order: $\mathrm{\tau_1 \rightarrow \tau_5 \rightarrow \tau_2 \rightarrow \tau_3 \rightarrow \tau_4}$.  

%due to switching between tasks that have less overlap between them. Instead, ordering the tasks in this order:  results in the least overhead.     

\parlabel{Cost Matrix.} The cost matrix $\mathrm{C_{n,n}}$ is an $\mathrm{n \times n}$ matrix, in which, each entry $\mathrm{c_{i,j}}$ denotes the additional cost of loading and executing task $\mathrm{\tau_j}$, given that the last executed task was $\mathrm{\tau_i}$. Note that the cost can be measured in terms of time or energy.   
%of switching tasks is proportional to the number of blocks on a task graph that are loaded and executed. In a multitask inference system involving $\mathrm{n}$ tasks, once the task graph has been formed, we compute the cost of switching, $\mathrm{c_{i,j}}$ from one task, $\mathrm{\tau_i}$ to the next task, $\mathrm{\tau_j}$ for all pairs of tasks. This results in a cost matrix, $\mathrm{C_{n,n}}$ of size $\mathrm{n \times n}$ as shown in Equation~\ref{eq:costmatrix}.        
\begin{equation}
\mathrm{C_{n,n}} = 
\begin{pmatrix}
0 & \mathrm{c_{1,2}} & \mathrm{c_{1,3}} & \cdots & \mathrm{c_{1,n}} \\
- & 0 & \mathrm{c_{2,3}} & \cdots & \mathrm{c_{2,n}} \\
- & - & 0 & \cdots & \mathrm{c_{3,n}} \\
\vdots  & \vdots  & \vdots  & \ddots & \vdots  \\
- & - & - & \cdots & 0 
\end{pmatrix}
\label{eq:costmatrix}
\end{equation}
The cost matrix is symmetric. Hence, only the upper or the lower triangular entries of the matrix are measured. These values are obtained empirically by measuring the time or energy overhead of switching between all pairs of tasks.

%, or analytically from the path length between the leaf nodes on the task graph. 

The cost matrix explains why task execution order matters. If all entries of the cost matrix were the same, the execution order of the tasks would not matter. This may only happen in extreme cases when tasks are too similar (i.e., they share all intermediate layers) or too different (i.e., they share nothing). The cost matrix is used to find the optimal execution order of the tasks.    

\subsection{Optimal Task Execution Order}
\label{sec:problemform}

%Given a task graph having $\mathrm{n}$ tasks and a task switching cost matrix, $\mathrm{C_{n,n}}$, our goal is to find an optimal ordering of the tasks so that the total cost of executing the task set is minimized.  

%https://www.tutorialspoint.com/design_and_analysis_of_algorithms/design_and_analysis_of_algorithms_np_hard_complete_classes.htm

Given a set of $\mathrm{n}$ tasks, $\mathrm{\tau=\{\tau_1,...,\tau_n \}}$ and cost matrix, $\mathrm{C_{n,n}}$, our goal is to find an optimal ordering of the tasks so that the total cost of executing the task set is minimized.  

\parlabel{Mathematical Formulation.} We define a binary variable $\mathrm{x_{ij}}$ to denote whether a task switching happens from $\mathrm{\tau_i}$ to $\mathrm{\tau_j}$:
  \begin{equation}
    \mathrm{x_{i,j}} =
    \begin{cases}
      1, & \text{if}\ \text{ a task switch happens from } \mathrm{\tau_i} \text{ to } \mathrm{\tau_j}\\
      0, & \text{otherwise}
    \end{cases}
  \end{equation}
The task ordering problem is formulated as the following integer linear programming problem:
\begin{equation*}
\begin{array}{lr@{}ll}
\text{minimize}     & \mathrm{\displaystyle\sum_{i=1}^n\sum_{j=1,j\neq i}^n}  & \mathrm{c_{i,j}x_{i,j}}      & \\
\text{subject to}   & \mathrm{\displaystyle\sum_{i=1,i\neq j}^n}              & \mathrm{x_{i,j}} = 1        & \mathrm{j = 1 ,\cdots, n}\\
                    & \mathrm{\displaystyle\sum_{j=1,j\neq i}^n}              & \mathrm{x_{i,j} = 1}        & \mathrm{i = 1 ,\cdots, n}\\
                    & \mathrm{\displaystyle\sum_{i\in Z}\sum_{j\neq i,j\in Z}} &  \mathrm{x_{i,j} \leq |Z| - 1} & \mathrm{\forall Z\subsetneq \{1 ,\dots, n\}, |Z|\leq 2}
\end{array}
\end{equation*}
Here, the objective function minimizes the overall task switching overhead for all tasks. The first two constraints ensure that tasks are executed only once. The last constraint ensures that there is no subset that can form a sub-tour and thus the final solution is a single execution order and not a union of smaller sub-orders~\cite{papadimitriou1998combinatorial}. 

\subsection{Inter-Task Dependencies}
\label{sec:tspvariants}

We have thus far discussed multitask inference scenarios where every task is executed in an orderly manner. In many real-world systems, however, there are additional constraints on these tasks that affect their execution decision. We categorize these into two broad classes: \emph{precedence} constraints and \emph{conditional} constraints.

\begin{figure}[!htb] 
    \centering
    %\vspace{-15pt}
    \subfloat[Precedence Constraint]
    {
        \includegraphics[width=1.64in]{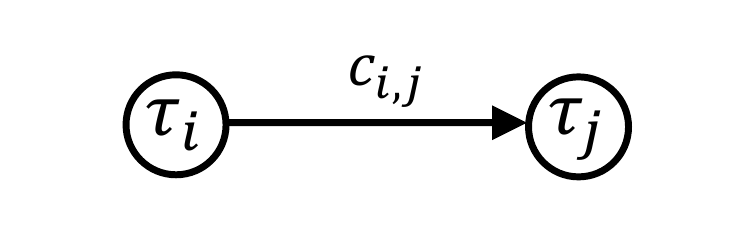}
    }%
    \subfloat[Conditional Constraint] 
    {
        \includegraphics[width=1.64in]{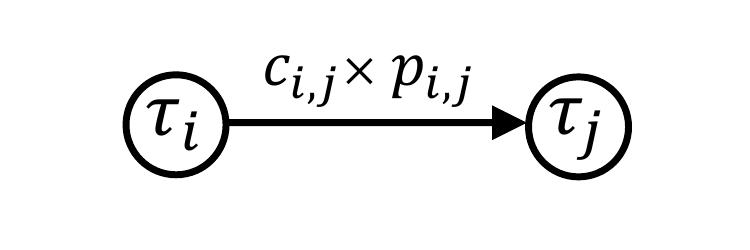}
    }
    \caption{Precedence and conditional constraints are expressed by directed edges. Unlike precedence constraints, conditional constraints affect the task switching cost.} 
    %\vspace{-15pt}
    \label{fig:algo_top_sort} 
\end{figure}

\parlabel{Precedence Constraints.} These constraints dictate that certain tasks (prerequisites) must be executed prior to some other tasks (dependents). These constraints are static. They are determined at the design time of the classifiers. We express these constraints using directed edges on a graph as shown in Figure~\ref{fig:algo_top_sort}(a) where each node, $\mathrm{\tau_i}$ denotes a task and each edge, $\mathrm{(\tau_i, \tau_j)}$ denotes a precedence constraint such that $\mathrm{\tau_i}$ must finish before $\mathrm{\tau_j}$ starts.

% \begin{figure}[!htb]
%     \includegraphics[height=1in]{figure/blank.png}%figure/algo/algo2_tsp_graph.pdf
%     \vspace{-5pt}
%     \caption{Precedence and conditional constraints are expressed by directed edges. Unlike precedence constraints, conditional constraints affect the task switching cost.}
%     \vspace{-10pt}
%     \label{fig:algo_top_sort}
% \end{figure} 

To account for the precedence constraints, we augment the optimization problem described in Section~\ref{sec:problemform} with additional constraints. We assume a given set of precedence constraints, $\mathrm{P}$ of tuples of tasks, $\mathrm{(\tau_i, \tau_j) \in P \subseteq |\tau| \times |\tau|}$, for which, the task, $\mathrm{\tau_j}$ must start after $\mathrm{\tau_i}$ finishes. For each task-pair, $\mathrm{(\tau_i, \tau_j)}$, we define the remaining execution time to finish $\mathrm{\tau_j}$, given that $\mathrm{\tau_i}$ has already finished, as $\mathrm{d_{i,j}}$. To formally incorporate precedence constraints, we define a new binary variable: 
  \begin{equation}
    \mathrm{s_{i,t}} =
    \begin{cases}
      1, & \text{if}\ \mathrm{\tau_i} \text{ starts by time } \mathrm{t}\\
      0, & \text{otherwise}
    \end{cases}
  \end{equation}
The following constraint ensures the inclusion of all precedence constraints:
\begin{equation}
\begin{array}{lll}
\mathrm{\displaystyle\sum_{t'\le t} s_{i,t'}} & \mathrm{\geq} &  \mathrm{\displaystyle\sum_{t'\le t + d_{i,j}} s_{j,t'}}    
\end{array}
\label{eq:pconstraint}
\end{equation}

\parlabel{Conditional Constraints.} These are a special type of precedence constraints where the decision to execute a dependent task depends on the outcome of a prerequisite task. These constraints manifest dynamically at runtime when a prerequisite task finishes and its inference result is available. 

%Conditional constraints are common in embedded systems where often a lightweight inference task is executed to determine whether a heavyweight inference task needs to run. For instance, voice assistant devices such as Alexa and Siri constantly run the wake-word detection task and wakes the rest of the inference modules only if a specific wake-word is detected.            

%The TSP with Conditional Constraint (TSPCC) is built on top of TSPPC. In real world scenarios, it is common that a bigger task that can produce higher-resolution results is executed conditionally based on the execution results of a smaller task. For example, an embedded system may first run a wake-word detection task to detect if there is a wake-up command from the user, e.g. "Hey Siri" for iPhone, and then decide whether or not to execute the task of fully analyzing the following audio clip. 

%It is worth noting that the result of the smaller wake-up task varies each time in real-world situation and thus our conditional constraint has a dynamic property which we can use a transfer probability to represent.

%To the best of our knowledge, the TSPCC is the first variant with tasks which are conditionally executed in a dynamic fashion.

Conditional constraints are also represented by directed edges. However, these constraints being dynamic, to accommodate their effect on the task switching cost (which in turn affects the task ordering), we utilize their probability of execution. We estimate this probability offline over a dataset by counting the fraction of the time a dependent task is executed after its prerequisite task finishes. We assume a given set of conditional constraints, $\mathrm{R}$ of triplets $\mathrm{(\tau_i, \tau_j, p_{i,j}) \in R \subseteq |\tau| \times |\tau| \times [0, 1]}$, where $\mathrm{(\tau_i, \tau_j) \in P}$ and $\mathrm{p_{i,j}}$ is the probability of executing $\mathrm{\tau_j}$ after $\mathrm{\tau_i}$ finishes. This probability is used to determine the expected cost of switching to a dependent task as shown in Figure~\ref{fig:algo_top_sort}.         

Since conditional constraints are a special type of precedence constraints, we include the same linear constraints as in Equation~\ref{eq:pconstraint} to account for them in the optimization problem. 

%The edge costs of the conditional constraint graph are utilized when solving the optimization problem which is described in the next section.    

% \begin{align}
%   \{(\tau_i, \tau_j)\} \subseteq P,\quad \forall (\tau_i,\tau_j,p_{ij})\in R
% \end{align}

%In Figure~\ref{fig:algo_top_sort}(b), each arc has a  weight which represents the transfer probability of the corresponding task pair. The probability will be used later to calculate the expected switch overhead for a given execution order. For arcs that do not have conditional constraint, we assign a probability of 1 to them. 

\subsection{Solving the Optimization Problem}

\parlabel{Brute-force Solver.} In extremely resource-constraint systems, we expect a small number of inference tasks. In such cases, a brute-force solver would suffice that generates all possible permutations of the tasks, discards the permutations that violate precedence constraints, and selects the best ordering that maximizes a fitness score. We define \emph{fitness score} for each permutation that does not violate the precedence constraints as the sum of task switching overheads: 
\begin{equation}
\begin{array}{ll}
\mathrm{f(\pi_1, \cdots , \pi_n)} &= \mathrm{\displaystyle\sum_{1 \le i < n} c_{\pi_i,\pi_{i+1}}}   
\end{array}
\label{eq:fit1}
\end{equation}
where, $\mathrm{\pi_i}$ refers to the task that executes at position $\mathrm{i}$. 

Specifically for conditional constraints, we adjust the fitness score to account for the probabilistic execution of dependent tasks by multiplying the probability to the switching cost:  
\begin{equation}
\begin{array}{ll}
\mathrm{f(\pi_1, \cdots , \pi_n)} &= \mathrm{\displaystyle\sum_{1 \le i < n} p_{\pi_i\pi_{i+1}}~c_{\pi_i,\pi_{i+1}}}
\end{array}
\label{eq:fit2}
\end{equation}

\parlabel{More Efficient Solver.} Although a brute-force solver is reasonably fast for a small number of tasks, the solver is repeatedly invoked during the task graph generation step -- once for each task graph as they are enumerated -- which as a whole takes significant time. Furthermore, in the future, the number of inference tasks running on low-resource systems could increase significantly as technology advances. Hence, we propose an efficient, scalable, genetic algorithm-based solver to solve the task ordering problem. The details of the algorithm are in the Appendix~\ref{sec:appendix1}.

\section{System Implementation}
\label{sec:systemdesign}

This section describes the hardware, the embedded system software, and the application development tool.  
\subsection{Hardware}

\begin{figure}[!htb] 
    \centering
    \vspace{-10pt}
    \subfloat[16-bit Custom]
    {
        \includegraphics[height = 1.6in]{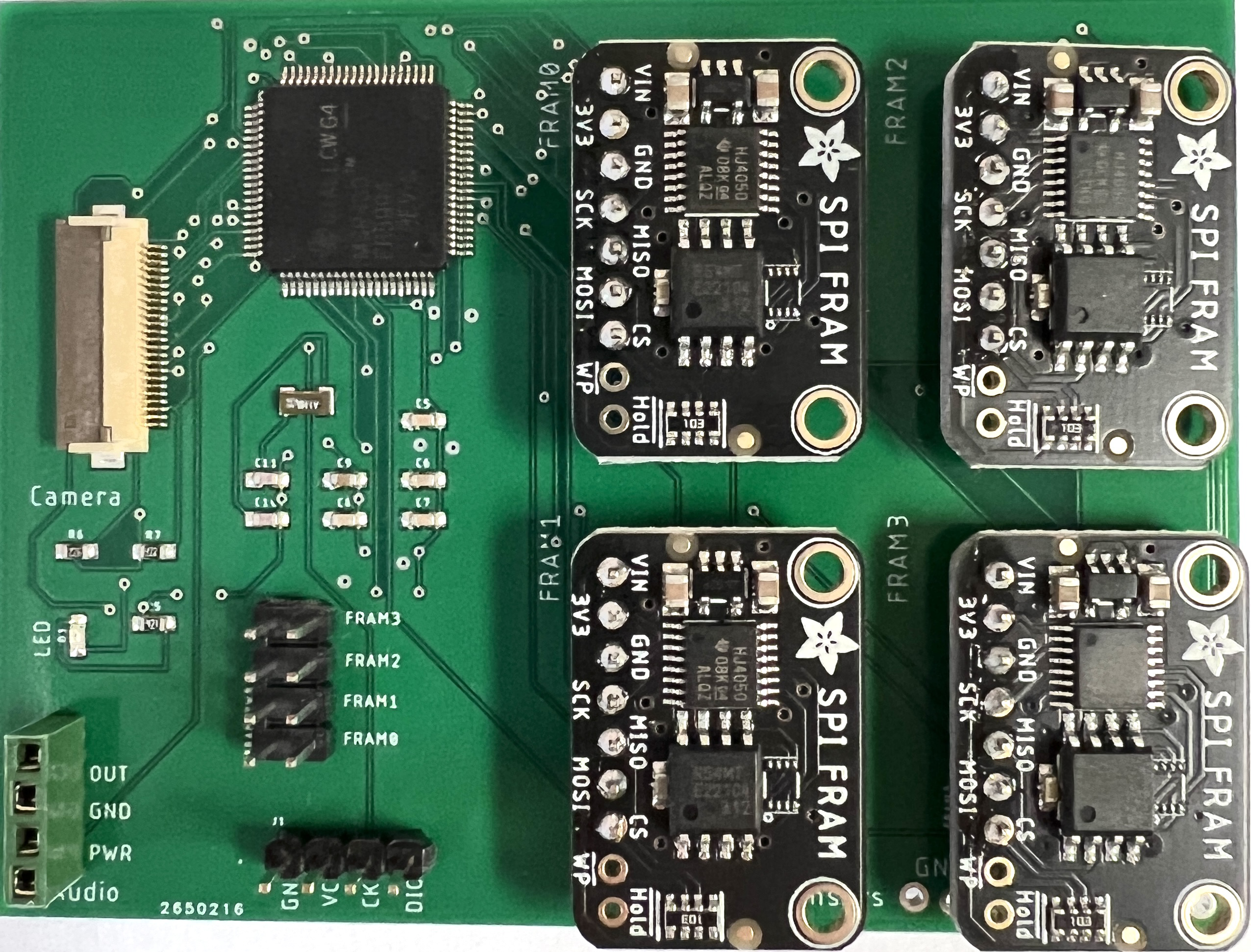}
    }
    \subfloat[32-bit Portenta] 
    {
        \includegraphics[height=1.6in]{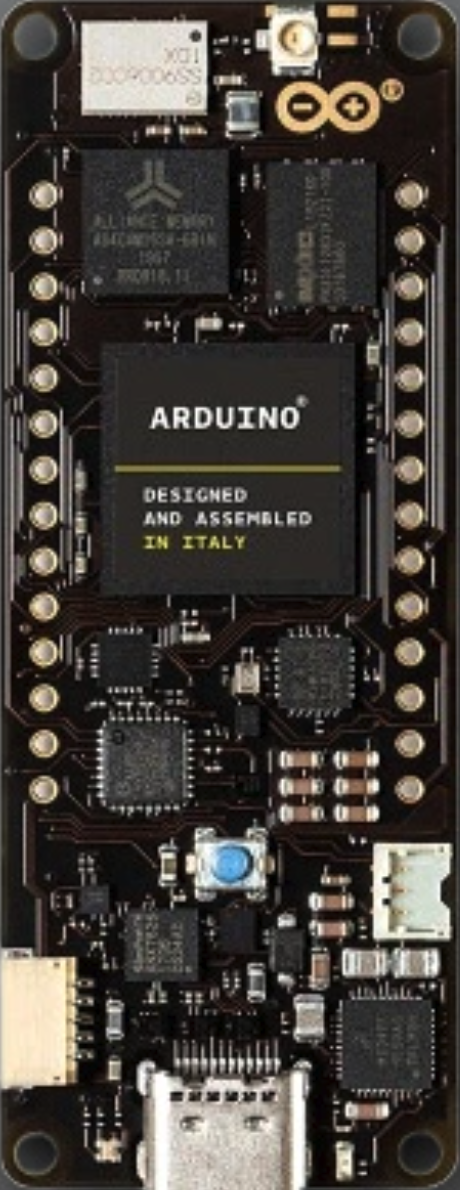}
    }
    \vspace{-5pt}
    \caption{Hardware platforms.} 
    \vspace{-10pt}
    \label{fig:pcb}
\end{figure}

\parlabel{Boards.} In order to implement and evaluate \sys, we use two hardware platforms: 1) a 16-bit TI MSP430FR5994-based custom-designed ultra-low-power system, and 2) a 32-bit ARM Cortex-M4/M7-based off-the-shelf Portenta H7. These two systems and their specifications are shown in Figure~\ref{fig:pcb} and Table~\ref{tab:hardwaretable}, respectively.    

\begin{table}[!htb]
    \centering
    \begin{tabular}{|l|l|l|}
    \hline
    \textbf{Platform} & \textbf{Custom} & \textbf{STM32H747}\\
    \hline
    CPU     &   MSP430FR5994   & ARM Cortex-M4/M7 \\
            &   16-bit, $\le$16MHz   & 32-bit, $\le$480MHz \\
    Memory  &   8KB SRAM    & 1MB SRAM\\
            &   512KB+2MB FRAM    & 2MB eFlash\\
    Power   &   1.8V - 3.6V     & 3.3V\\
            &   118 uA/MHz (active) & 100 mA \\
    \hline
    \end{tabular}
    \caption{Hardware specification.}
    \vspace{-20pt}
    \label{tab:hardwaretable}
\end{table}

The rationale behind using two different platforms is to demonstrate \sys's performance as we vary the CPU (16-bit vs. 32-bit), external memory type (FRAM vs. flash), and application (audio vs. image). Since off-the-shelf 16-bit MCUs have very limited on-board FRAM, we design a custom PCB that contains a 16-bit MSP430FR5994 MCU and expands its non-volatile memory from 512KB to 2MB+ by provisioning extra slots to connect up to four 512KB external FRAMs.         
For a fair comparison with YONO~\cite{yono}, we choose the same platform (STM32H747) in 32-bit experiments, enforce similar restrictions on the memory usage, and use the same core (M7).

%\noindent \textbf{System memory.} As shown in Figure~\ref{fig:pcb}, the platform have an array of four FRAM chips. We use off-the-shelf FRAM breakout board\cite{exfram} developed by Adafruit and each breakout board has 512KB of memory. Though our current design can only store up to 2MB of external models, this FRAM array can be easily extended if needed. We use MSP430FR5994 as the on-board MCU. It is a ultra-low power processor and has build-in 256KB of FRAM and 8KB of SRAM memory. 

%\vspace{2pt}
%\noindent \textbf{Peripheral sensors.} 

\parlabel{Sensors.} Both systems have interfaces to connect a camera, a microphone, and an inertial measurement unit (IMU). We use an ultra-low-power camera (HM01B0~\cite{hm01b0}) that consumes <2mW at QVGA 30FPS for image sensing and a MAX4466~\cite{max4466} microphone for audio sensing.

%The MCU has a dedicated hardware accelerator module called Low Energy Accelerator (LEA) which can greatly accelerate neural model inference if used properly.

\subsection{Embedded System Software}

%capuchin: TFlite -> model -> C header + library -> C code;  

All neural network inference-related software that run on the target device are referred to as the embedded system software. One of the challenges to this has been the portability of \sys across different platforms such as MSP430FR5994 and STM32H747. To streamline this, we follow a two-step process: 

First, we develop a Python-based tool to convert a TensorFlow model to a C header file containing the weights and the architecture of the neural network. Then a task-specific C header file for each task is generated. All such task-specific C header files are combined (as dictated by the task graph) to generate a single compact header file containing network weights, the common network architecture, and task graph information.    

Second, the header file generated above is combined with platform-specific C implementation of neural network modules. A complete C program is auto-generated, which executes the neural networks in the desired order. The code is  cross-compiled to produce executable binary files for the target device. The tool supports sequential TensorFlow models and has a C library that implements dense layers, convolution, maxpooling, flattening, dropout, and leaky ReLU operations. 
  
\subsection{Application Development Tool}

We provide a Python-based application development tool that takes a dataset and the tasks as the input, and produces a task graph and task execution order as the output. There are three steps to the process:

First, neural network instances are created and trained on the given dataset to obtain task-specific neural networks using Tensor Flow. These Tensor Flow models are auto-converted to C files for the target platform. Execution time of each layer of these networks are profiled by running them on the target platform. These measurements are used later in the task graph generation and selection process.       

% Second, task affinity vs. overhead trade-off analysis is performed for a given range of normalized model size budgets. 

%\bnote{We use the tradeoff point (shown in Figure~\ref{fig:algo1_tradeoff}) as our model size budget.}

Second, variety score vs. overhead tradeoff analysis is performed for the full range of model size budgets. Branch point parameter is set to 3. All task graphs within the budget range are generated, and their variety scores and execution overheads are recorded in a file. Task graph that balances variety score and execution overhead is selected by default. However, the developer at this point may choose to use any task graph from the file that meets their application and system requirements.   

Third, All tasks are retrained on the input dataset using a multitask learning algorithm~\cite{vandenhende2019branched}.  

%The switching cost matrix is estimated. Optimal ordering of tasks is determined.   

\section{Evaluation}
\label{sec:eval}

This section describes two sets of experiments: evaluation of the proposed algorithms and end-to-end system performance.   

\subsection{Experimental Setup}

\parlabel{Dataset and Network Architecture.} We use the datasets and network architectures used in recent multitask inference literature for low-resource systems~\cite{nwv,yono,nws}. Table~\ref{table:network} provides a summary. The network architecture shown on the table (rightmost column) is used as the common network architecture in \sys and each task on a dataset corresponds to recognizing one class. All datasets have 10 tasks, except for HHAR which has 6. We use 80\% of the data for training and 20\% for testing.

%However, as previous literature is all about heterogeneous tasks where each task is a multi-class classification task but our system is for homogeneous tasks where all tasks use the same input, we have to  convert each $n$-class classification dataset into $n$ binary classification tasks. $\mathrhm{n}$ is 6 for HHAR dataset and 10 for all other datasets. 

%We use standard datasets and network architecture, listed on Table~\ref{table:network}, which were used in recent multitask inference literature for low-resource systems~\cite{nwv,yono,nws}. We use the same for a fair comparison. However, as previous literature is all about heterogeneous tasks where each task is a multi-class classification task but our system is for homogeneous tasks where all tasks use the same input, we have to  convert each $n$-class classification dataset into $n$ binary classification tasks.  $\mathrhm{n}$ is 6 for HHAR dataset and 10 for all other datasets. We use 80\% of the data for training and 20\% for testing.

% To implement multitask learners, instead of one $\mathrm{n}$-way classifier, we design $\mathrhm{n}$ binary classifiers. $\mathrhm{n}$ is 6 for HHAR dataset and 10 for all other datasets. We use 80\% of the data for training and 20\% for testing.

\begin{table}[!htb]
\begin{tabular}{|l|l|l|}
%\toprule
\hline
\textbf{Modality}                  & \textbf{Dataset} & \textbf{Architecture}  \\ \hline
\multirow[t]{6}{*}{Image} & MNIST~\cite{lecun1998gradient}  & LeNet-5~\cite{lecun1998gradient} \\
                  & F-MNIST~\cite{xiao2017fashion} & LeNet-5~\cite{lecun1998gradient} \\
                  & CIFAR-10~\cite{krizhevsky2009learning} & DeepIoT~\cite{yao2017deepiot}  \\
                  & SVHN~\cite{netzer2011reading} & Neuro.Zero~\cite{lee2019neuro} \\
                  & GTSRB~\cite{stallkamp2011german} & LeNet-4~\cite{lecun1995comparison} \\ \hline
\multirow[t]{3}{*}{Audio} & GSC-v2~\cite{warden2018speech} & KWS~\cite{chen2014small} \\
                  & ESC~\cite{piczak2015esc} & Mixup-CNN~\cite{zhang2018deep} \\
                  & US8K~\cite{salamon2014dataset} & TSCNN-DS~\cite{su2019environment}\\ \hline
IMU                  & HAAR~\cite{stisen2015smart} & DeepSense~\cite{yao2017deepsense}  \\ 
%\bottomrule
\hline
\end{tabular}
\caption{Datasets and Network Architectures.}
\label{table:network}
\vspace{-20pt}
\end{table}

\parlabel{Baselines for Comparison.} We use four baselines for comparison: YONO~\cite{yono}, NWV~\cite{nwv}, NWS~\cite{nws} and Vanilla. The first three~\cite{yono, nwv, nws} are the state-of-the-art. Vanilla refers to independently trained classifiers running sequentially on the system. We use NWV and NWS in both 16-bit and 32-bit experiments. Since their source codes are not available, we use our own implementation and cross-check with their reported results to ensure that they are consistent with ours. We use YONO only in 32-bit experiments. Since our 32-bit hardware platform is identical to YONO's, we use the reported results from their paper~\cite{yono}.      

%whose source codes are not available. 

%We use our own implementation of NWV and NWS. Since YONO and \sys use the same hardware platform and same  

%Our system as the \emph{\sys}.          

\parlabel{Evaluation Platforms.} We use the two platforms described earlier (Section~\ref{sec:systemdesign}), i.e., a 16-bit MSP430FR5994 and a 32-bit STM32H747. Data samples are pre-loaded into the non-volatile memory from where they are read and executed. 
We measure the time and energy consumption by connecting a 100$\Omega$ resistor in series with the board and using Analog Discovery~\cite{discovery} to measure the voltage across the resistor.

\parlabel{Code Release.} All software and hardware designs will be made open source if this work is accepted.

%%%%%%%%%%%%%%%%%%%%%%%%%%%%%%%%%%%%%%%%%%%%%%%%%
%%%%%%%%%%%%%%%%%%%%%%%%%%%%%%%%%%%%%%%%%%%%%%%%%
\subsection{Algorithm and Parameter Analysis}

This section describes how parameters such as  branch points, variety-overhead trade-off points, and dependency constraints affect the algorithms proposed in Sections~\ref{sec:taskgraphgeneration} and ~\ref{sec:taskorder}.

%We use a 10-layer network which occasionally required us to extend some of the networks in Table~\ref{table:network} as they have less than 10 layers. 

\begin{figure}[!thb] 
    \centering
    \subfloat[Variety Score]
    {
        \includegraphics[width = 0.49\textwidth]{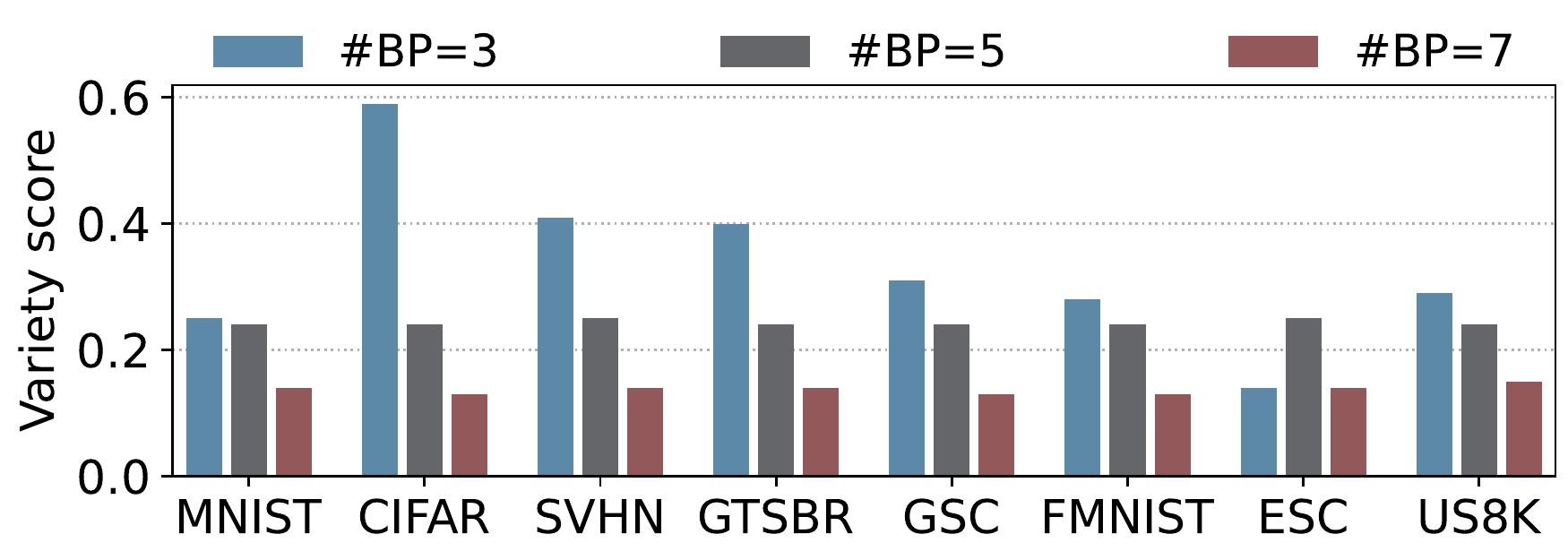}
    }
    
    \subfloat[Execution Cost] 
    {
        \includegraphics[width = 0.49\textwidth]{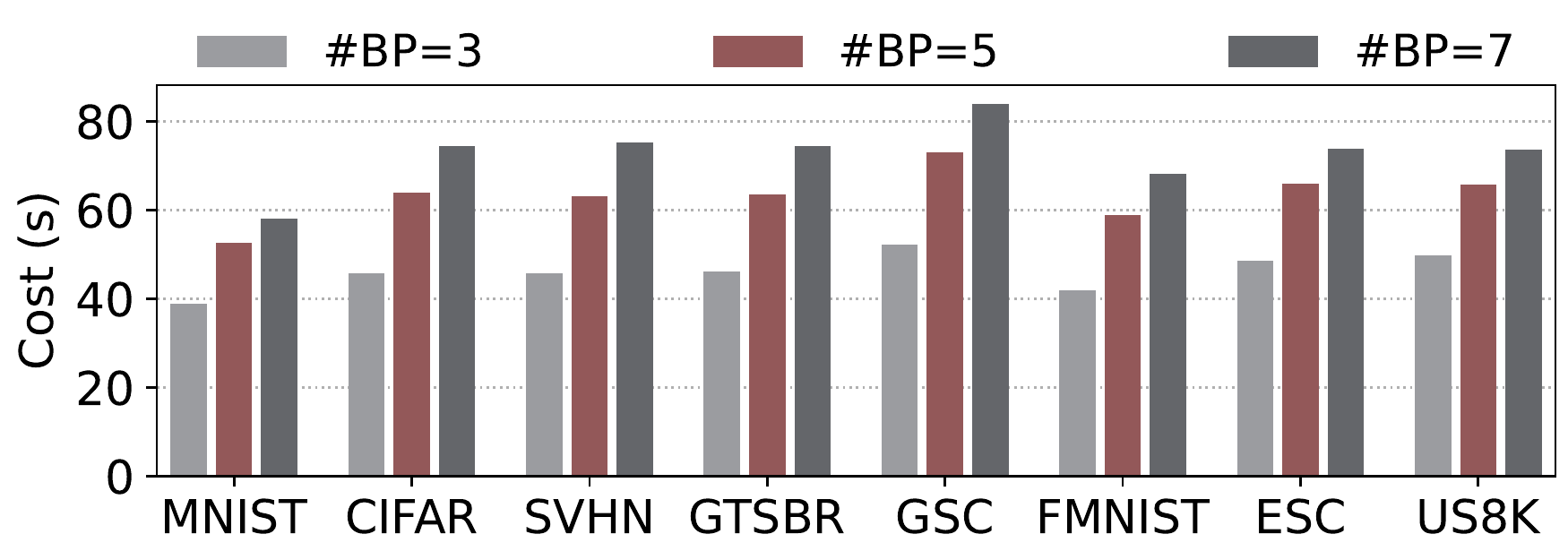}
    }
    \caption{Effect of the number of branch points.} 
    \label{fig:eval_bpnumber} 
    \vspace{-15pt}
\end{figure}

\parlabel{Effect of Branch Point Choice.} We evaluate the effect of branch points by performing a sensitivity analysis of variety and execution cost. We use execution time as cost. We vary the number of branch points, $\mathrm{BP = \{3, 5, 7\}}$. Results are shown in Figure~\ref{fig:eval_bpnumber}. We observe that more branch points improve variety score (lower is better) but worsen the overhead. This is because more branch points decompose and groups tasks at a finer granularity which causes more tasks to branch out at deeper layers and thus decreases task switching efficiency.

\begin{figure}[!thbt] 
    \centering
    \subfloat[Variety Score]
    {
        \includegraphics[width = 0.49\textwidth]{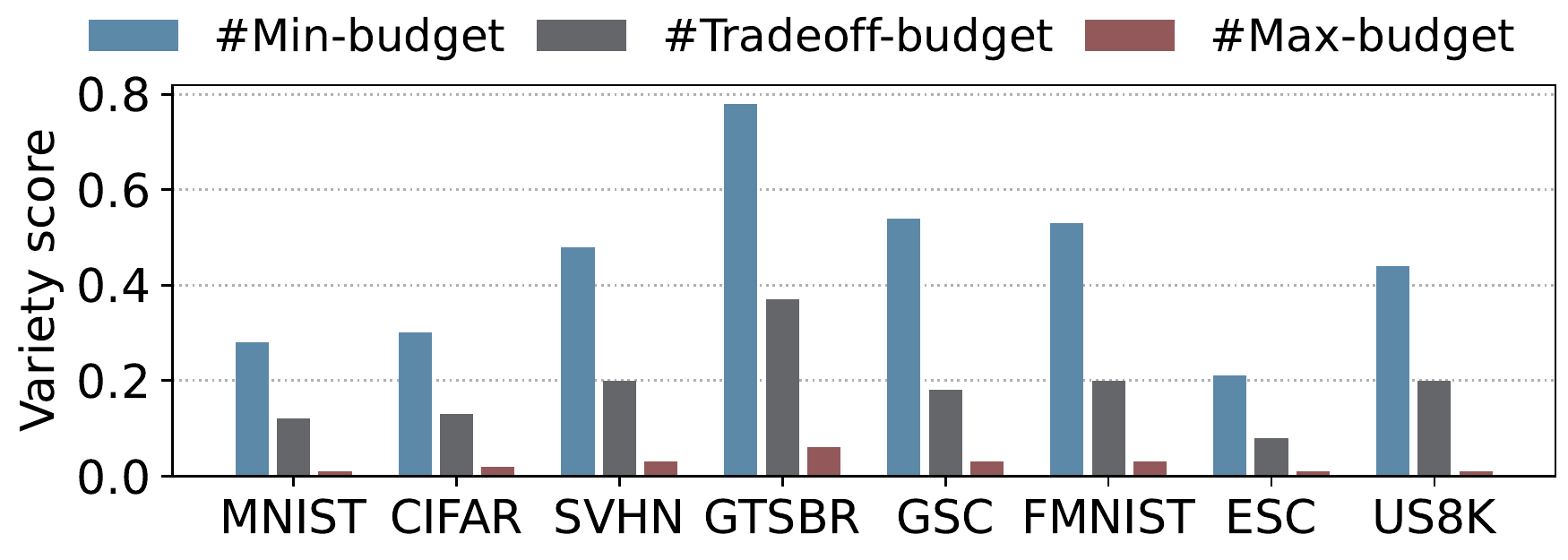}
    }
    
    \subfloat[Execution Cost] 
    {
        \includegraphics[width = 0.49\textwidth]{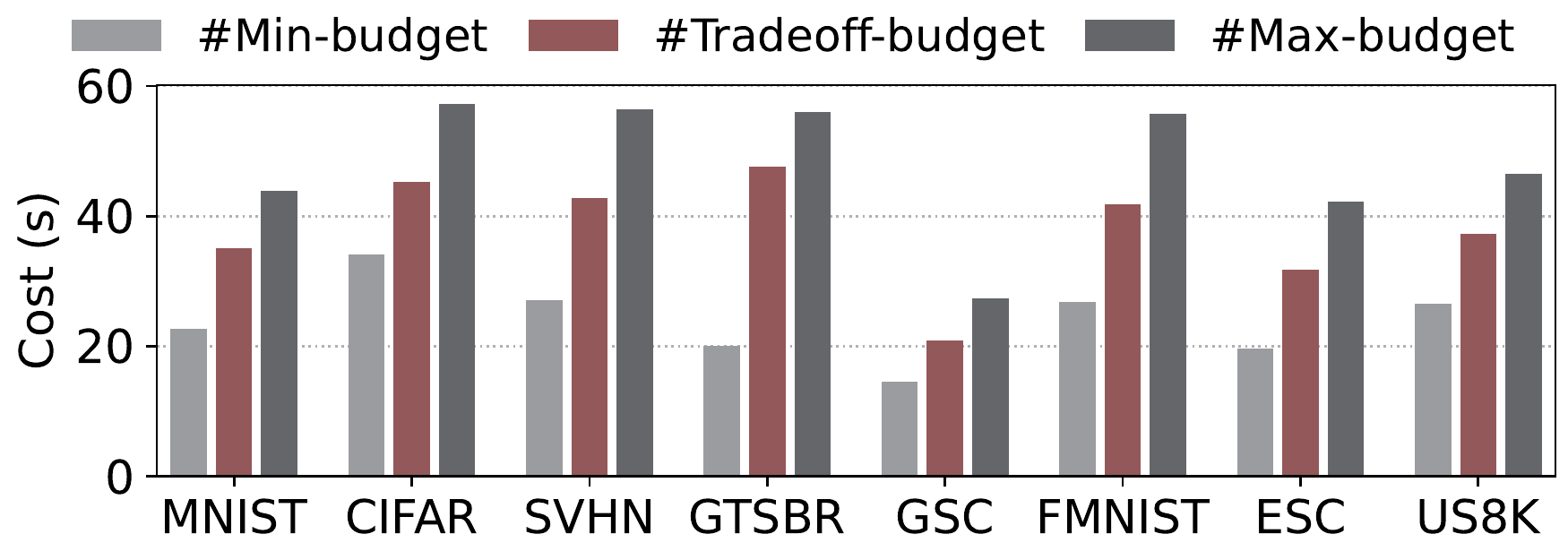}
    }
    \vspace{-10pt}
    \caption{Variety score vs. execution cost tradeoff.}
    \vspace{-5pt}
    \label{fig:eval_bplocation} 
\end{figure} 

\parlabel{Variety Score vs. Execution Cost Tradeoff.} Figure~\ref{fig:eval_bplocation} shows the tradeoff between variety score and execution cost for each dataset. We compare three network size budgets: two extreme cases of minimum and maximum budget, and a tradeoff budget where variety and cost trend lines intersect. We observe that low budget favors execution cost, high budget favors variety, and the trade-off budget balances the variety score and execution cost.

\begin{table}[!htb]
\resizebox{\linewidth}{!}{
    \centering
    \begin{tabular}{|l|l|l|l|l|}
    \hline
    %\textbf{TSP} & \textbf{Dataset} & \textbf{\# of} & \textbf{\# of} & \textbf{\# of} & \textbf{Ground} & \textbf{Our} \\
    \textbf{Variant} & \textbf{Dataset} & \textbf{Node/ Pre/Cnd} & \textbf{Optimal} & \textbf{\sys} \\\hline
    
    \multirow[t]{3}{*}{Regular} & FIVE & 5/0/0 & 19 & 19 \\
    & P01 & 15/0/0 & 291 & 291 \\
    & GR17 & 17/0/0 & 2085 & 2085 \\ \hline
    
    \multirow[t]{5}{*}{Precedence} & ESC07 & 9/6/0 & 2125 & 2125 \\
    &ESC11 & 13/3/0 & 2075 & 2075 \\
    % &ESC12 & 14/7/0 & 1675 & 1675 \\
    % &br.17.10 & 17/10/0 &55 & 55 \\
    &br17.12 & 17/12/0 & 55 & 55 \\ \hline
    
    \multirow[t]{3}{*}{Conditional} & ESC07 & 9/6/3 & 982 & 982 \\
    &ESC11 & 13/3/3 & 1901 & 2000 \\
    &ESC12 & 14/7/3 & 1398 & 1423 \\ \hline
    %&br.17.10 & 17 & 10 & 3 &/ & 55 \\
    %&br17.12 & 17 & 12 &3 & / &  55\\ \hline
    \end{tabular}
    }
    \caption{Evaluation of genetic algorithm for task ordering.}
    \vspace{-20pt}
    \label{table:genetic_algorithm}
\end{table}
%%%%%%%%%%%%%%%%%%%%%%%%%%%%%%%%%%%%%%%%%%%%%%%%%
\parlabel{Performance of Genetic Algorithm.}  
We evaluate the performance of the genetic algorithm for task ordering. For this, we use a popular public dataset for Traveling Salesperson Problem (TSP) called the TSPLIB~\cite{tsplib} and repurpose it for task ordering problem. This dataset already contains test cases that have precedence constraints. To include conditional constraints, we add weights to the edges of the graph. Table~\ref{table:genetic_algorithm} compares our results with the ground truth for all three cases of task ordering problem. Our result is identical to the ground truth for all cases except for a few conditional constraint cases with 5\% deviation.

\subsection{Comparison with Baseline Solutions}

\begin{figure}[!htb] 
    \centering
    \subfloat[16-bit MSP430FR5994]
    {
        \includegraphics[width = 0.49\textwidth]{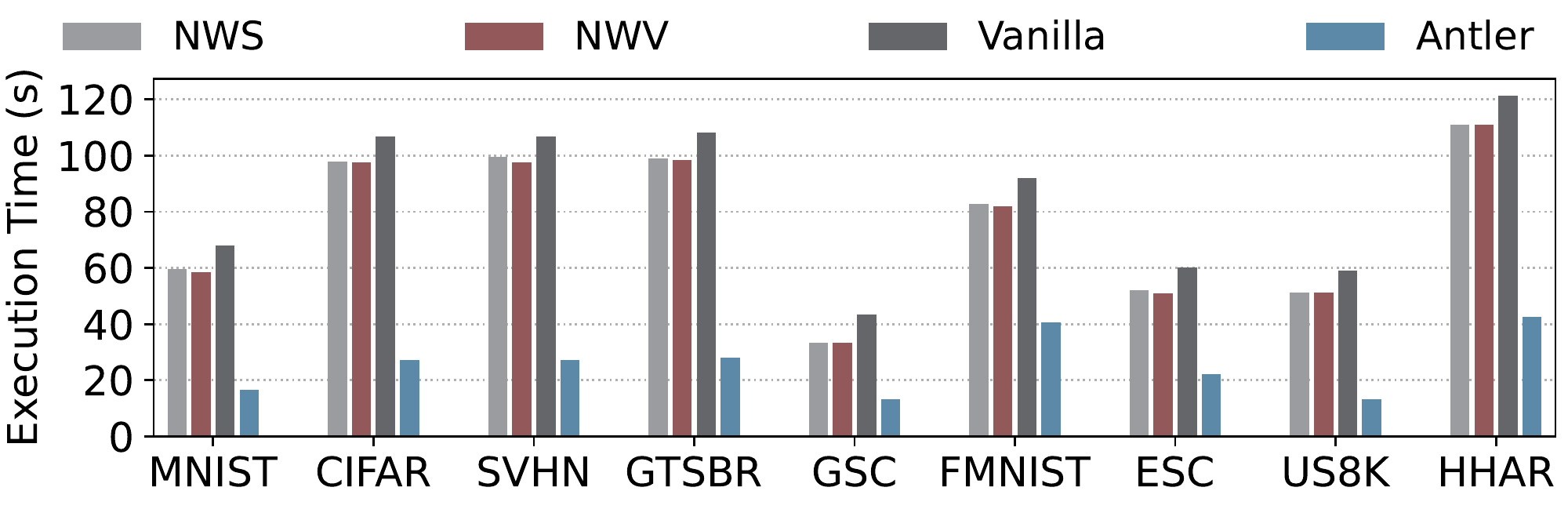}
    }
    
    \subfloat[32-bit STM32H747] 
    {
        \includegraphics[width = 0.49\textwidth]{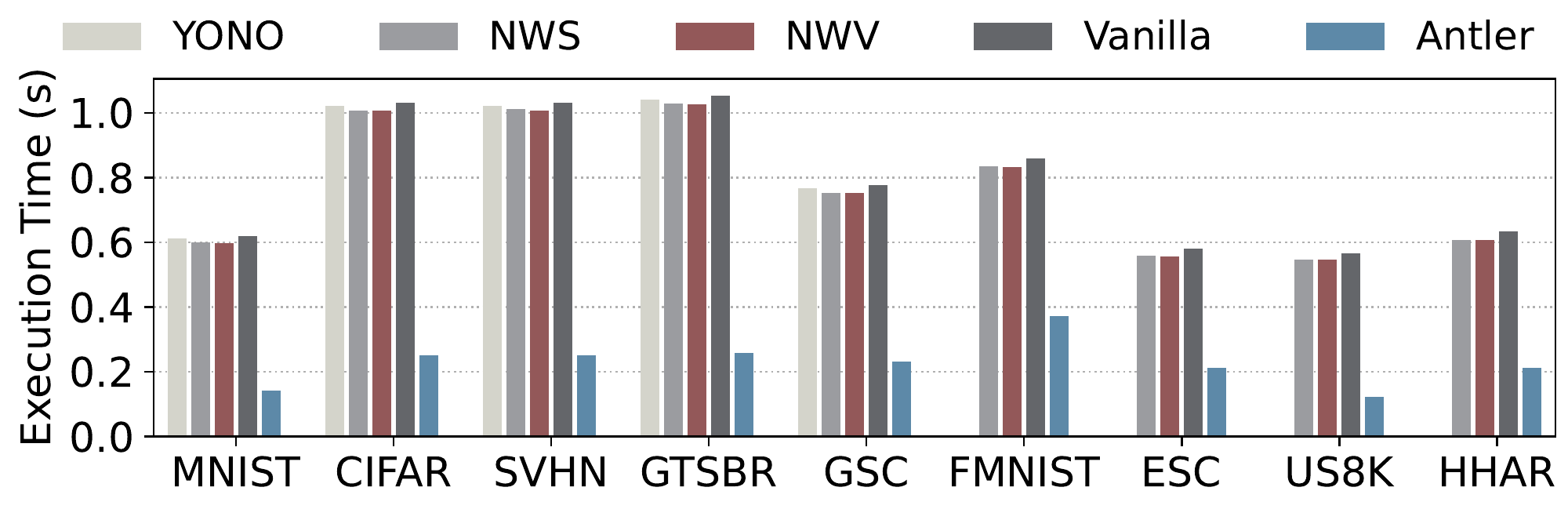}
    }
    \vspace{-5pt}
    \caption{Comparison of execution time.} 
    \vspace{-15pt}
    \label{fig:eval_timeoverhead} 
\end{figure}

\begin{figure}[!htb] 
    \centering
    \subfloat[16-bit MSP430FR5994]
    {
        \includegraphics[width = 0.49\textwidth]{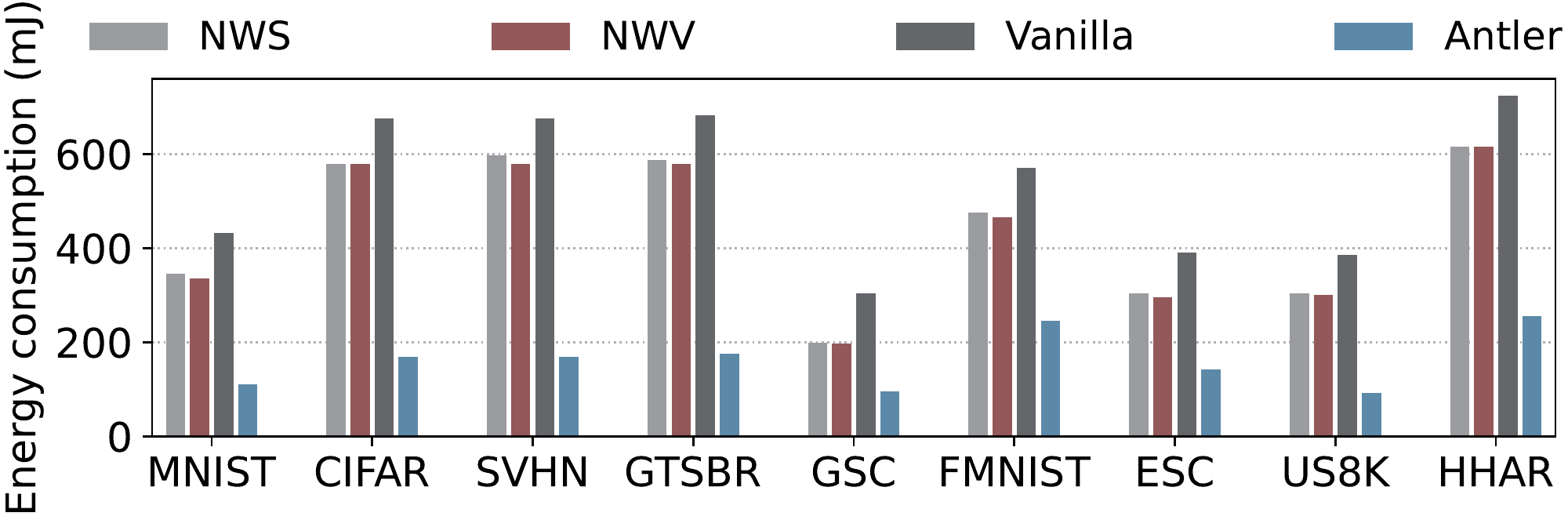}
    }
    
    \subfloat[32-bit STM32H747] 
    {
        \includegraphics[width = 0.49\textwidth]{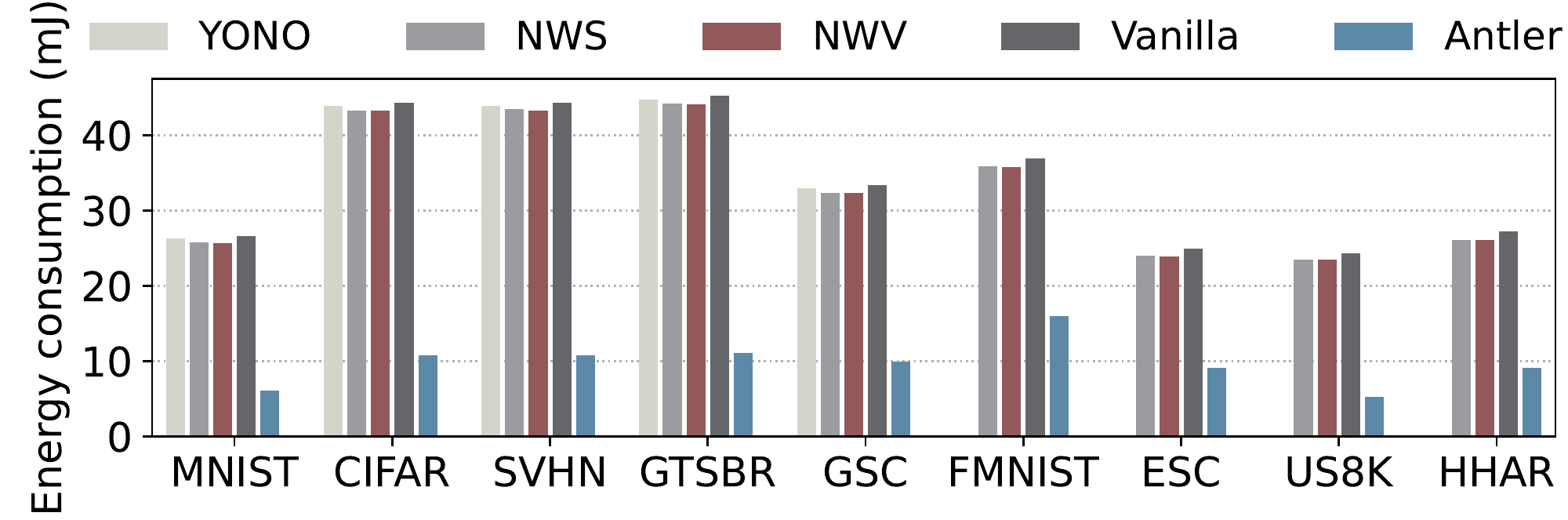}
    }
    \vspace{-5pt}
    \caption{Comparison of energy consumption.} 
    \vspace{-15pt}
    \label{fig:eval_energyoverhead} 
\end{figure}

%%%%%%%%%%%%%%%%%%%%%%%%%%%%%%%%%%%%%%%%%%%%%%%%%
%%%%%%%%%%%%%%%%%%%%%%%%%%%%%%%%%%%%%%%%%%%%%%%%%
\parlabel{Execution Time and Energy Cost.} We compare the execution time and energy consumption of \sys against the baselines in Figure~\ref{fig:eval_timeoverhead} and Figure~\ref{fig:eval_energyoverhead}, respectively. The Y-axis shows the total execution time (or energy) to execute all tasks for an input. We report results for both 16-bit and 32-bit systems.

%Since we use reported results of YONO, we only compare with YONO on 32-bit Portenta  (the same hardware used in YONO) for fairness.

We observe that while the general trend remains the same in both systems, the execution time on STM32H747 is 100X faster. On both systems, \sys's execution time is the lowest. This is because \sys leverages the similarity of tasks and reuses the intermediate results to reduce the execution time by 2.3X -- 4.6X which baseline solutions do not. Even though NWV and YONO perform complete in-memory inferences and have zero switching overhead, they fall short of \sys as the cost of repeatedly executing shared subtasks is higher, especially when it involves convolution layers. We observe similar pattern in energy consumption. Overall, \sys saves 56\% -- 78\% energy compared to the baselines.

\begin{figure}[!htb] 
    \centering
    \subfloat[time breakdown]
    {
        \includegraphics[width = 0.22\textwidth]{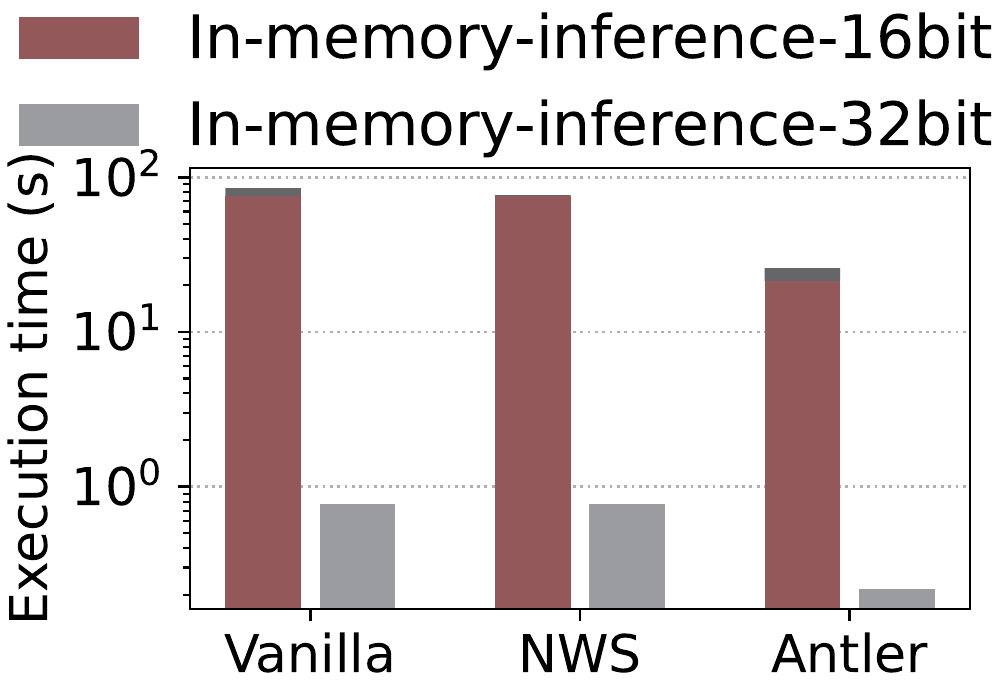}
    }
    \subfloat[energy breakdown] 
    {
        \includegraphics[width = 0.22\textwidth]{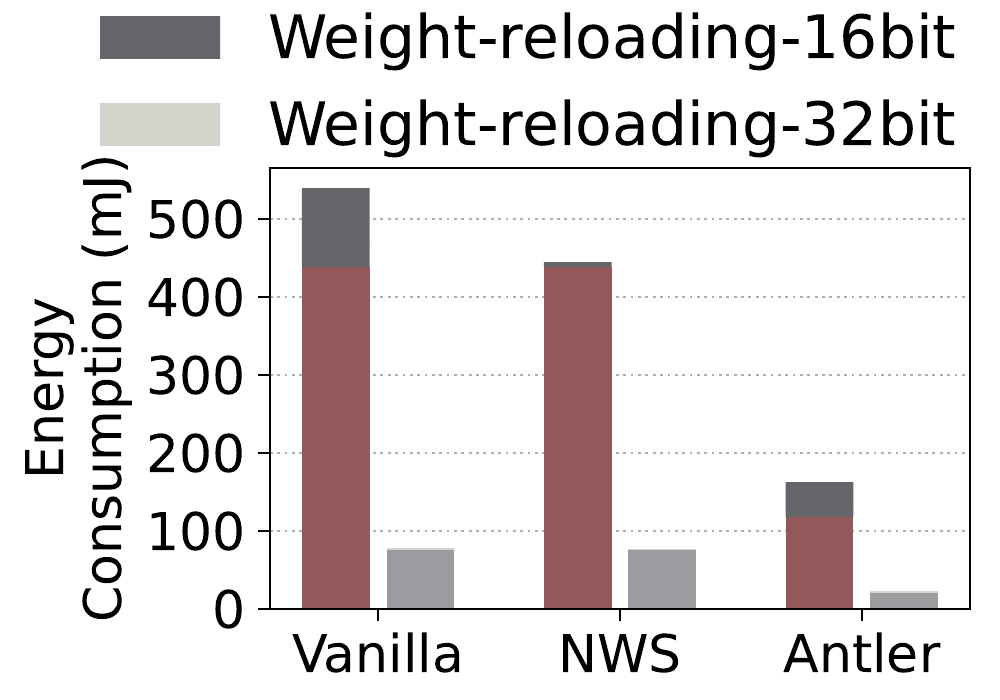}
    }
    \caption{Time and energy cost breakdown.} 
    \vspace{-15pt}
    \label{fig:eval_breakdown} 
\end{figure}

%%%%%%%%%%%%%%%%%%%%%%%%%%%%%%%%%%%%%%%%%%%%%%%%%
%%%%%%%%%%%%%%%%%%%%%%%%%%%%%%%%%%%%%%%%%%%%%%%%%
\parlabel{Breakdown of Time and Energy Overhead.}
We breakdown the total time and energy cost into two parts: inference-only cost that corresponds to in-memory execution of the networks and switching overhead that corresponds to loading weights from external memory. We compare \sys with Vanilla and NWS since the other two (NWV and YONO) do not use external memory and thus have no switching cost. Results are shown in Figure~\ref{fig:eval_breakdown}. 

The Y-values in Figures~\ref{fig:eval_breakdown}(a) and ~\ref{fig:eval_breakdown}(b) are averaged over all datasets. We observe that 32-bit STM32H747 has very little weight reloading overhead (the striped area on top of each bar is almost invisible) for both time and energy breakdown.
The time and energy cost related to weight reloading in NWS is also negligible as it only has around 7\% of the total weights stored in external memory.
\sys's time and energy cost related to weight-reloading is 54\%-56\% less than Vanilla. 

% We observe that 32-bit Portenta incurs very low switching overhead and its inference time  much less than that of 16-bit MSP430.
% However, their energy overhead is somewhat comparable because the RP2040 board is more power-hungry and its power consumption is 7X of MSP430's. 

\begin{figure}[!htb] 
    \centering
    \includegraphics[width=\linewidth]{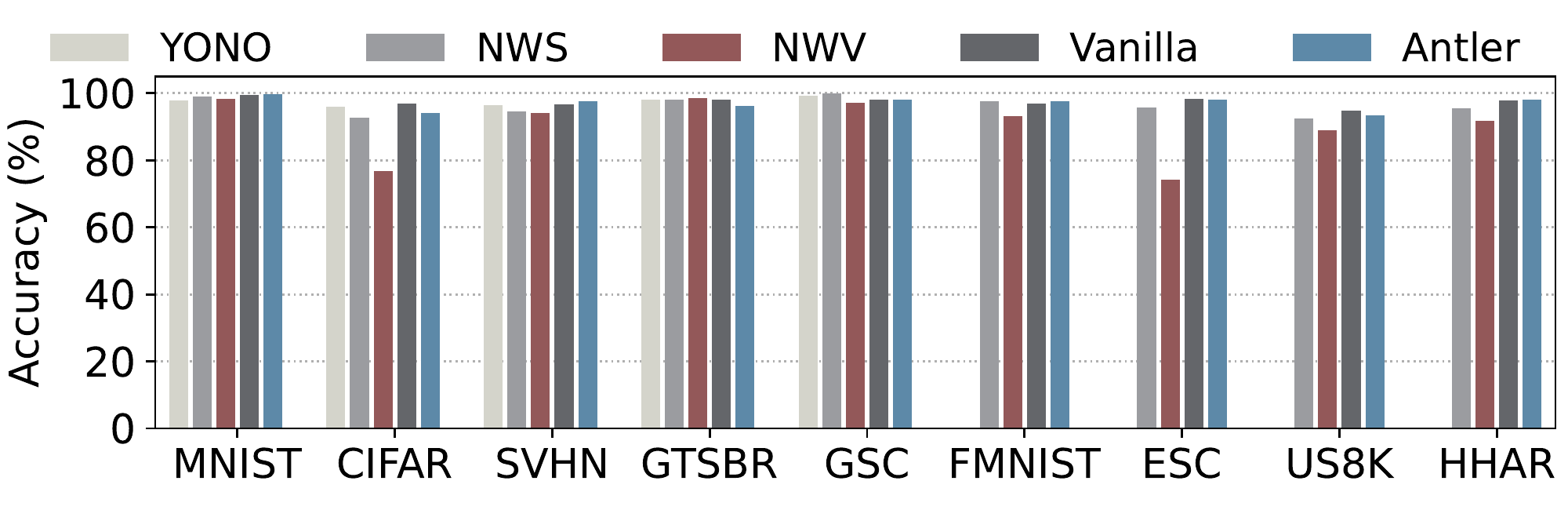}
    \vspace{-12pt}
    \caption{Comparison of inference accuracy. }
    \vspace{-7pt}
    \label{fig:eval_accuracy}
\end{figure}

\parlabel{Inference Accuracy.} We compare the inference accuracy of all systems in Figure~\ref{fig:eval_accuracy}. The accuracy is averaged over all tasks. \sys's inference accuracy is similar to YONO, NWS, and Vanilla within a margin of $\pm 3\%$ deviation. Recall that \sys's target is to reduce the time and memory cost of inference while achieving a high accuracy. In this case, all classifiers show reasonably high accuracy of over 90\%, except for NWV whose accuracy does not scale with the number of tasks. YONO does not use the later five datasets and thus its accuracy could not be included in Figure~\ref{fig:eval_accuracy} for those datasets.           

\parlabel{Memory Efficiency.}
We measure the total memory consumption of all tasks for each baseline and summarize in Table~\ref{table:memory}. We observe that \sys consumes more memory than NWS, NWV and YONO. This is because NWV and YONO perform complete in-memory execution and thus they are limited by the size of the RAM. Unlike them, \sys and NWS are able to utilize external memory and put no hard restrictions on the total size of the tasks. \sys consumes significantly less memory than Vanilla since \sys reduces memory consumption by exploiting the task affinity.  

\begin{table}[!htb]
\resizebox{\linewidth}{!}{
\begin{tabular}{|l|lllll|}
\hline
 \textbf{System} & Vanilla & Antler & NWS & NWV & YONO \\ \hline
 \textbf{Memory (KB)} & 1328 & 587 & 213 & 140 & 114   \\ \hline
\end{tabular}
}
\caption{Comparison of memory consumption.}
\vspace{-15pt}
\label{table:memory}
\end{table}

%%%%%%%%%%%%%%%%%%%%%%%%%%%%%%%%%%%%%%%%%%%%%%%%%
\section{Real-World Deployment}

We deploy \sys in two real-world multitask learning scenarios that involve audio and image classification tasks.

%respectively.

%application scenarios. We develop two multitask learners that perform audio and image classification, respectively.    

%All of the previous works\cite{yono,nws,nwv} only evaluated their proposed approach on datasets, which is not enough to show the performance in the wild. We evaluate the performance of \sys\ in the wild by implementing real-world audio-based and image-based experiments. 

\begin{figure}[!htb] 
    \centering
    \subfloat[Multitask Audio Inference System]
    {
        \includegraphics[width = 0.5\linewidth]{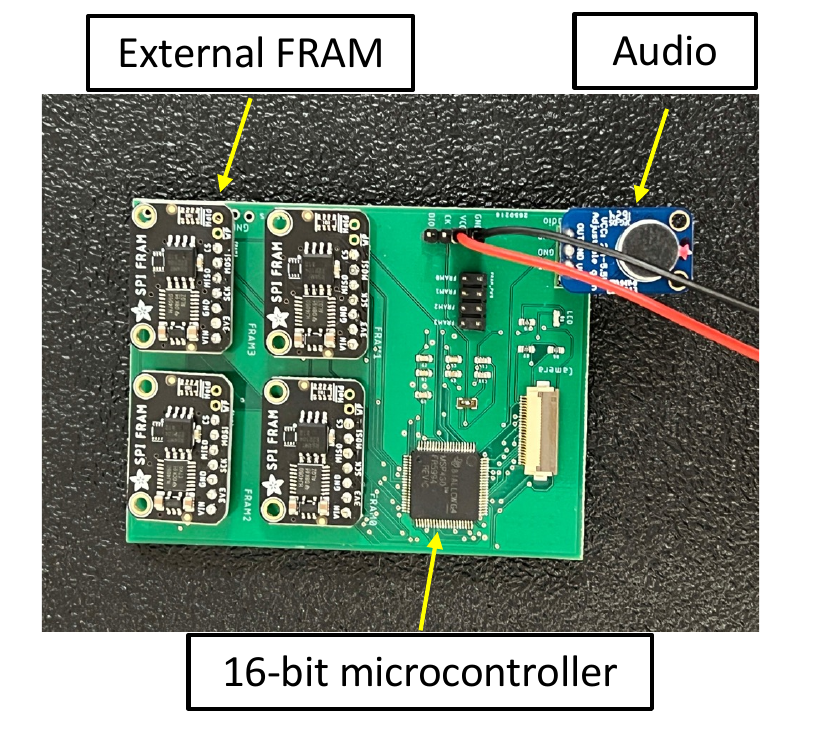}
        \includegraphics[width = 0.5\linewidth]{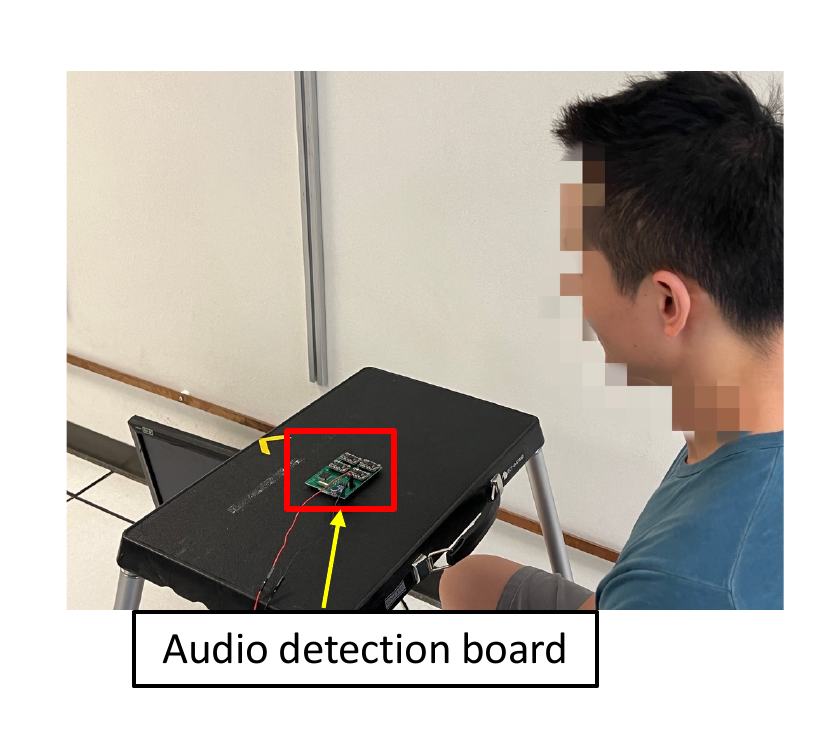}
    }
    
    \subfloat[Multitask Image Inference System] 
    {
        \includegraphics[width = 0.5\linewidth]{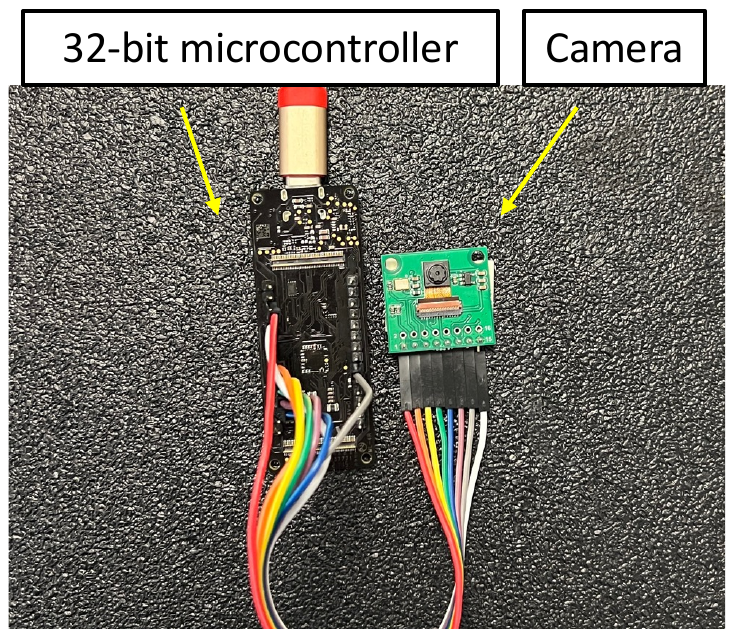}
        \includegraphics[width = 0.5\linewidth]{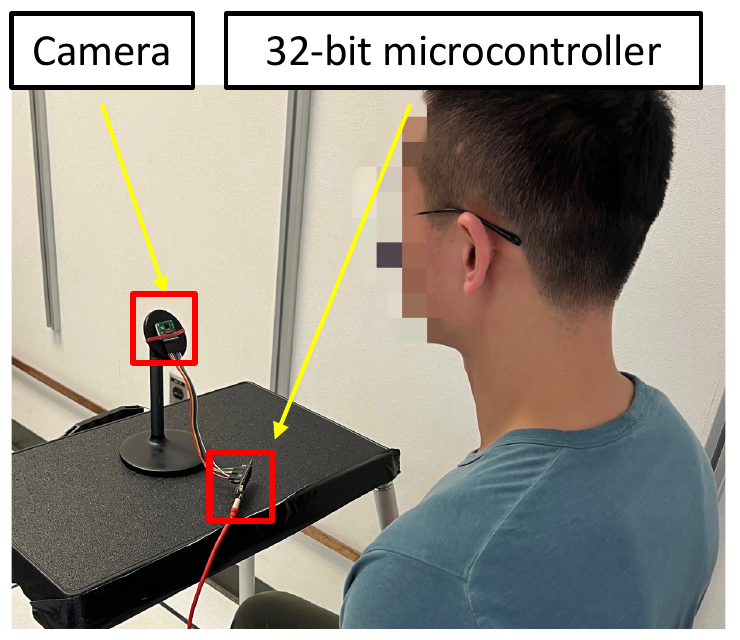}
    }
    \caption{Deployment Setup.} 
    \vspace{-15pt}
    \label{fig:setup_wild} 
\end{figure} 

%\subsection{Experimental Setup}
\subsection{Multitask Audio Inference System}

\parlabel{Inference Tasks.}
We implement five audio-based tasks: 1) a speaker presence detection task ($\tau_0$) which detects if there is human voice in the audio, 2) a command detection task ($\tau_1$) which detects eleven commands \{yes, no, up, down, go, stop, left, right, on, off, Alexa\}, 3) a speaker identification task ($\tau_2$) which identifies who is speaking (five speakers), 4) an emotion classification task ($\tau_3$) which classifies the audio into three emotions \{positive, negative, neutral\}, and 5) a distance classification task ($\tau_4$) which tells whether the speaker is close to or far from the device. 

%The eleven commands we use are:  The three emotion types are: .

\parlabel{System Setup.} We use the 16-bit custom MSP430FR5994~\cite{msp430fr5994} to conduct the audio-based experiments as shown in Figure~\ref{fig:setup_wild}(a). 
Audio signal is sampled at 2KHz and converted to a feature map having a window-length of 128ms and a stride of 64ms after performing the FFT. 

% \begin{figure}[!htb]
%     \centering
%     \includegraphics[width=\linewidth]{figure/ppt/deployment.pdf}
%     \caption{Multitask inference graphs}
%     \label{fig:taskgraphdeployment}
%     %\vspace{-20pt}
% \end{figure}

\begin{figure}[!htb] 
    \centering
    %\vspace{-15pt}
    \subfloat[Audio Inference Graph]
    {
        \includegraphics[width = 0.45\linewidth]{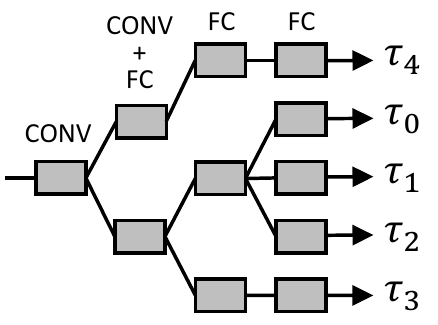}
    }%
    \subfloat[Image Inference Graph] 
    {
        \includegraphics[width = 0.45\linewidth]{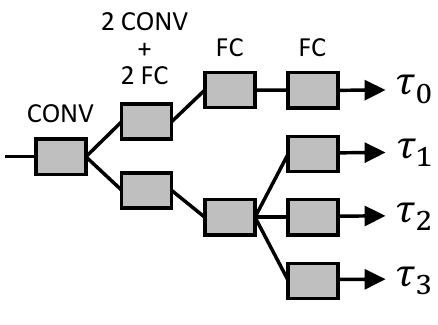}
    }
    \caption{Multitask inference graphs.} 
    \vspace{-15pt}
    \label{fig:wildgraphs} 
\end{figure} 

\parlabel{Data Collection and Network Training.} Five volunteers (four male and one female) participate in this experiment. We have followed Institutional Review Board (IRB) approved protocol to conduct this study. We collect 15 samples for each task from each volunteer. We use 80\% data for training and 20\% for testing. We design a 5-layer CNN having 2 convolutional and 3 dense layers and pre-train it on~\cite{warden2018speech} prior to training on our own dataset. We use 3 branch points and the tradeoff budget for the task graph. 

\subsection{Multitask Image Inference System}

\parlabel{Inference Tasks.}
We implement four image classification tasks: a human presence detection task ($\tau_0$) which detects human faces in an image, a mask detection task ($\tau_1$) which detects if the person is wearing a mask, a person identification task ($\tau_2$) which recognizes a person's face (5 volunteers), and an emotion recognizer ($\tau_3$) which classifies three emotions as in the audio inference system. 

\parlabel{System Setup.}
We use off-the-shelf 32-bit STM32H747 H7 as shown in Figure~\ref{fig:setup_wild}(b). 
Images are taken with a HM01B0 camera module and has the dimensions of 64$\times$64 pixels.  

\parlabel{Data Collection and Network Training.} Data collection and network training processes are identical to the audio inference system except for the neural network which is a 7-layer CNN having 3 convolutional and 4 dense layers and is pre-trained on ~\cite{LFW}.

\subsection{Evaluation Results}

%We evaluate the memory usage, inference accuracy, and  time and energy overhead of the two systems. We also show the task decomposition and grouping details. 

\parlabel{Task Decomposition and Grouping.} Figure~\ref{fig:wildgraphs} shows task graphs for both applications. There are 4 blocks in each task graph (for 3 branch points). One of the blocks (second block) contain multiple layers. This is unlike task graphs observed earlier in Section~\ref{sec:eval} where deeper layers are lumped into the same block. Overall, having more layers lumped in earlier blocks decreases execution cost but may decrease accuracy as well. \sys finds a trade-off point between the two to optimize both accuracy and cost.           

%decomposition and grouping results. We can see that most layers are decomposed into Block$_1$ for both audio and image applications. However, we observe that for those dataset based experiments in  \S\ref{sec:eval}, most layers are decomposed into Block$_2$. Overall, assigning more layers to earlier blocks decreases cost but may deteriorate accuracy and \sys\ finds out the sweet trade-off point between the two.

\parlabel{Task Dependency and Ordering.} We include a precedence constraint in the image inference system that the presence detection task ($\tau_0$) must be executed before any other task. Additionally, in the audio inference system, we make presence detection a conditional constraint such that other tasks are executed at 80\% probability. The ordering of tasks in audio and image inference systems are: $\mathrm{\tau_0 \rightarrow \tau_3 \rightarrow \tau_4 \rightarrow \tau_2 \rightarrow \tau_1}$ and $\mathrm{\tau_0 \rightarrow \tau_3 \rightarrow \tau_1 \rightarrow \tau_2}$, respectively, and they are just one of several orderings that yield the best performance for these tasks.

%In \S \ref{sec:taskorder}, we discuss three different TSP problems, including general TSP, TSPPC and TSPCC. For general TSP, we just assume there is no dependency between tasks and evaluate it on both applications. 
%For TSPPC, we evaluate it on the image-based application and impose a dependency constraint which requires that all other tasks be executed only after the presence detection task $\tau_0$ is done. 
%For TSPCC, we evaluate it on the audio-based application and except for the above dependency constraint we further impose a conditional constraint which requires that only when the presence detection task $\tau_0$ detects a human face, other tasks will be executed and otherwise other tasks will be skipped. We assume the skipping probability is 0.8. 

\begin{figure}[!htb] 
    \centering
    \vspace{-5pt}
    \subfloat[Execution Time]
    {
        \includegraphics[width = 0.24\textwidth]{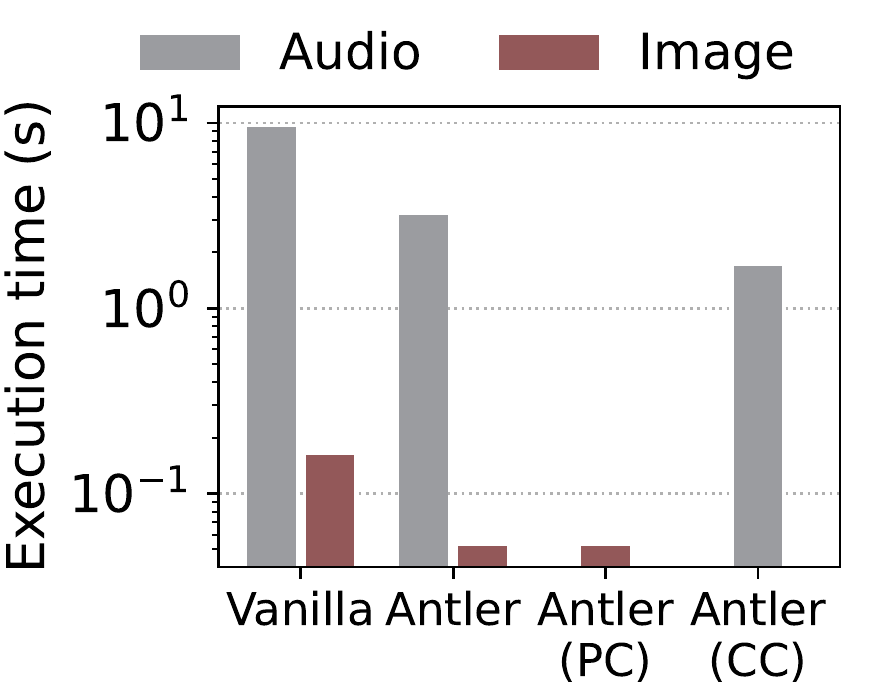}
    }
    \subfloat[Energy Consumption] 
    {
        \includegraphics[width = 0.24\textwidth]{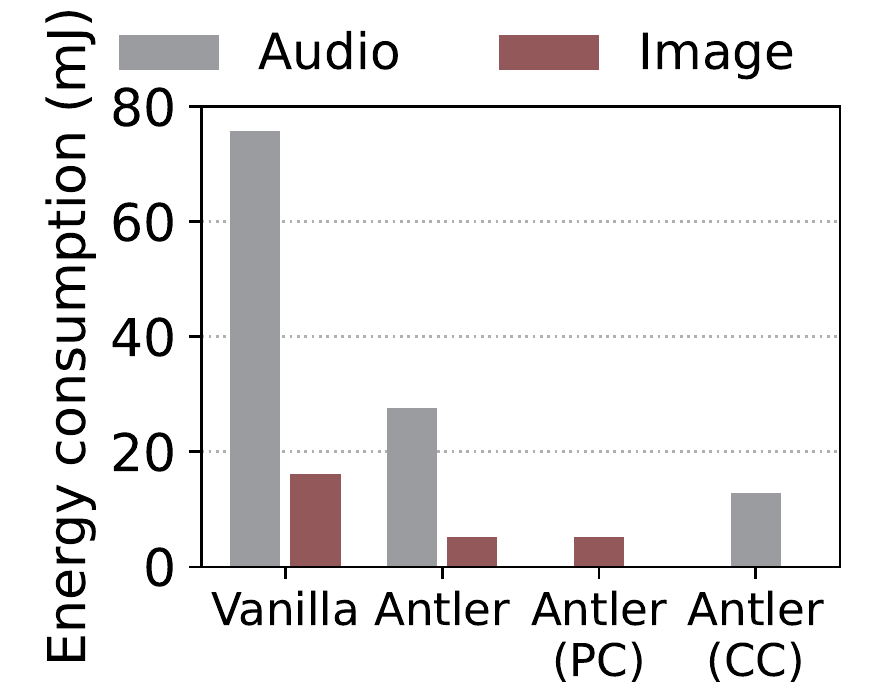}
    }
    \caption{Execution time and energy consumption.} 
    \vspace{-5pt}
    \label{fig:eval_overhead_wild} 
\end{figure}

\parlabel{Inference Time and Energy Consumption.}
We evaluate \sys's execution time and energy consumption for three cases: \sys having no constraints, \sys-PC having precedence constraints, and \sys-CC having conditional constraints, and compare their performance with the Vanilla system. Figure~\ref{fig:eval_overhead_wild} shows that \sys yields 2.7X -- 3.1X reduction in time and energy costs and the results are consistent across both systems. \sys-PC has the same overhead as \sys because there are only four tasks and the execution order with precedence constraint is already in the optimal ordering. For \sys-CC, overhead decreases as tasks are skipped occasionally based on the conditional probability.          

%on three TSP variants and compare it to Vanilla baseline. Results are shown in Figure~\ref{fig:eval_overhead_wild}. We can see that \sys\ has an 2.7X - 3.1X overhead reduction compared to Vanilla baseline with regard to both time and energy. The results are consistent on both audio and image based experiments. 
%For the TSPPC variant, it has the same overhead as the general version because we only have four tasks and the execution order with precedence constraint is already in the optimal ordering list. For the TSPCC variant, we can see that the overhead is decreased because tasks are skipped with a probability which decreases the weighted overall overhead.  

\begin{figure}[!htb] 
    \centering
    \vspace{-15pt}
    \subfloat[Audio Inference System]
    {
        \includegraphics[width = 0.24\textwidth]{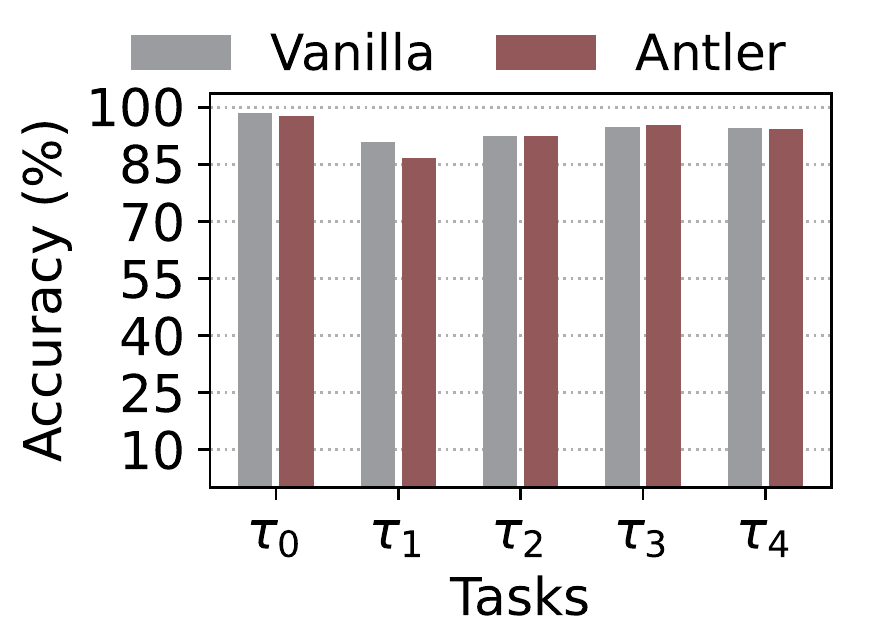}
    }
    \subfloat[Image Inference System] 
    {
        \includegraphics[width = 0.24\textwidth]{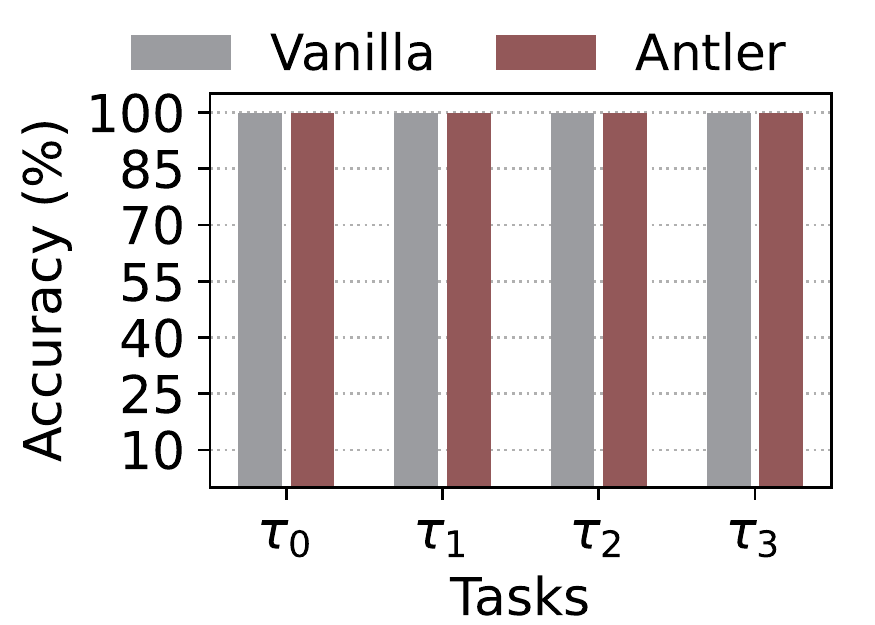}
    }
    \caption{Inference accuracy of audio and image classifiers.} 
    \vspace{-5pt}
    \label{fig:eval_wild_accuracy} 
\end{figure}

\parlabel{Inference Accuracy and Memory Usage.} 
Figure~\ref{fig:eval_wild_accuracy} shows the average accuracy of all tasks for both systems. We observe that in the audio inference system, except for the command detection task, all tasks have over 90\% accuracy. This is because the command detection task is the hardest of these tasks with eleven class labels. The accuracy of both \sys and Vanilla are very similar within an average deviation of $\pm 1\%$. 

%Besides, we can see that Vanilla baseline only has an visible advantage on the command detection task and for all other tasks all methods have similar results. This validates that Vanilla baseline is only preferred when each task individually is hard. 

The memory usage of both systems are shown in Table~\ref{table:memoryusage}. We observe that the memory usage of \sys is approximately half of Vanilla's, which is consistent with earlier results from the dataset-driven experiments in Section~\ref{sec:eval}. 

\begin{table}[!thb]
    \begin{tabular}{|l|lll|}
    \hline
    \textbf{System}                  &  & Vanilla & Antler \\ \hline
    \multirow{2}{*}{\textbf{Memory (KB)}} & Audio & 397 & 202 \\ \cline{2-4} 
                               & Image & 445 & 222 \\ \hline
    \end{tabular}
    \caption{Memory usage.}
    \vspace{-30pt}
    \label{table:memoryusage}
\end{table}

\section{Related Work}

%Speed, energy-efficiency, and accuracy are conflicting goals when comes to neural network inference on embedded systems. Any algorithmic technique to achieve one or more these goals inherently is a constrained optimization problem. In optimizing multitask inference, a common control knob in the literature is the network size since a larger network tends to show higher accuracy, but incurs higher time and energy costs. Hence, a vast majority of works focus on network compression.

\parlabel{Single Network Compression.} This class of algorithmic techniques refer to approaches that take one DNN at a time and compress it down to a desired size by employing a wide variety of methods such as knowledge distillation~\cite{chen2017learning}, low-rank factorization~\cite{sainath2013low}, pruning~\cite{guo2016dynamic}, quantization~\cite{han2015deep}, compression with structured matrices~\cite{cheng2015exploration}, network binarization~\cite{rastegari2016xnor}, and hashing~\cite{chen2015compressing}. The disadvantages of this technique are: first, there is no cross-DNN knowledge sharing or joint compression that trains multiple DNNs together; second, since each DNN is compressed individually using different compression methods, this technique is not scalable and they do not have the advantages of multitask learning; third, a significantly compressed DNN does not run nearly as significantly faster since most parameters are pruned in the dense layers while convolutional layers consume most computation time~\cite{han2015deep}.

There are DNN compression methods that do not yield much benefit in terms of memory usage since the assignments of weights to connections have to be stored additionally. This includes soft weight sharing approaches~\cite{ullrich2017soft} such as the Dirichlet process~\cite{roth2018bayesian}, k-means clustering~\cite{han2015deep}, and quantization~\cite{koksal2001weight}.  

\parlabel{Multiple Networks Compression.} This class of algorithmic techniques compress multiple DNNs together; e.g., PackNet~\cite{mallya2018packnet} compresses multiple DNNs to a single DNN with iterative pruning of redundant parameters in DNNs to remove weights that can be used by other tasks. The number of DNNs that can participate in the process, however, is limited when free weights fall short as the number of DNNs increase while a single network is maintained. ~\cite{chou2018merging} proposes a technique that merges DNNs by integrating convolutional layers. However, their technique works for two networks only and it requires layer alignment for merging. Learn-them-all~\cite{kaiser2017one} trains a single network to deal with many tasks simultaneously. However, choice of a suitable architecture is generally hard for learning all the tasks apriori. Besides, this requires a large training data from different types and sources which is a tedious task. 

There exists multiple studies on sharing weights among a set of DNNs, e.g., MultiTask Zipping~\cite{he2018multi} combines DNNs for cross-model compression with a layer-wise neuron sharing; Sub-Network Routing~\cite{ma2019snr} modularizes the shared layers into multiple layers of sub-networks;  Cross-stitch Networks~\cite{misra2016cross} apply weight sharing~\cite{duong2015low} after pooling and fully-connected layers of two DNNs.  The scope and methods of weight sharing in these works are limited by the choice of network architecture and task type.

% MMoE~\cite{ma2018modeling} learns to model task relationships from data by sharing the sub-models and weights across tasks; Tensor factorization~\cite{yang2016deep} divides each set of parameters into shared and task-specific parts;

\parlabel{Most Relevant Works.} The three most relevant state-of-the-art systems to \sys are NWV~\cite{nwv}, NWS~\cite{nws}, and YONO~\cite{yono}. NWV and YONO propose complete in-memory execution of DNNs on memory-constrained systems. NWS extends NWV by allowing some of the high-significance weights into the flash memory to increase the accuracy of NWV. NWS essentially points out that complete in-memory packing and execution of DNNs in MCUs is not accurate. All three approaches fail to leverage the affinity among tasks and their dependencies, and thus repeatedly execute overlapping common subtasks that significant increases the time and energy cost of multitask inference which is avoided by \sys.      

\section{Conclusion}

We envision a future where a wide variety of ultra-low-power sensing and inference systems will sense and classify every aspect of our personal and physical world. To realize this vision, we need to significantly lower the time and energy cost of running multiple neural networks on low-resource systems while ensuring that their application-level performance does not degrade. To achieve this goal, we propose \sys, which is the first system that exploits the similarity among a set of machine learning tasks to identify overlapping substructures in them that is combined and executed in an optimal order to reduce the execution time by 2.3X -- 4.6X and the energy overhead 56\% -- 78\% when compared to the state-of-the-art multitask learners for low-resource systems. 

\section*{Appendix}
\label{sec:appendix1}

%\parlabel{NP Completeness Proof.} 
\subsection{NP Completeness}

We prove that task ordering problem on an arbitrary task graph is NP-Complete. First, we show that the task ordering problem belongs to NP following a corresponding proof for the traveling salesperson problem (TSP)~\cite{korte2011combinatorial}. Since a task can switch to any task, a tour that contains each task exactly once can be constructed. The total cost of the edges of the tour is the sum of switching cost, $\mathrm{c_{i,j}}$ corresponding to the edge $\mathrm{(\tau_i, \tau_j)}$ on the tour. Finally, we determine if the cost is minimum. This is completed in polynomial time. Therefore, the task ordering problem is in NP.   

Second, we show that the task ordering problem is NP-hard by reducing an instance of a Hamiltonian cycle problem~\cite{korte2011combinatorial} to it. We take an instance of Hamiltonian cycle, $\mathrm{G(V, E)}$. From this, we construct an instance of task ordering problem. We construct a complete graph $\mathrm{G'(V, E')}$, where $\mathrm{E' = \{(u, v) | u, v \in V, i \ne j\}}$. Note that $\mathrm{G'}$ is not a task graph but rather a complete graph whose nodes are the tasks from the given task graph and edges are the cost of switching between tasks. We define a cost function as: 
  \begin{equation}
    \mathrm{\gamma(u, v)} =
    \begin{cases}
      0, & \text{if}\ \mathrm{(u, v) \in E} \\
      1, & \text{otherwise}
    \end{cases}
  \end{equation}
Using the cost function above we can argue that if a Hamiltonian cycle exists in $\mathrm{G}$, that cycle will have a cost of 0 in $\mathrm{G'}$ by construction. In other words, if $\mathrm{G}$ has a Hamiltonian cycle, we have an ordering of tasks of 0 overhead.

Conversely, we assume that $\mathrm{G'}$ has a tour (i.e., an ordering of tasks) of cost at most 0. Since edges in $\mathrm{E'}$ are 0 or 1, each edge on the tour (i.e., each task switching overhead on the chosen ordering of tasks) must be of cost 0 as the cost of the tour is 0. Therefore, the tour contains only edges in $\mathrm{E}$. 

This proves that $\mathrm{G}$ has a Hamiltonian cycle if and only if $\mathrm{G'}$ has an ordering of tasks of at most 0 overhead. 

%\parlabel{Genetic Algorithm Solver.} 

\subsection{Genetic Algorithm Solver}

For the general case, especially when the number of tasks is too large for the brute-force solver, we propose a genetic algorithm to solve the optimal task ordering problem. Genetic algorithm is an evolutionary search method which provides optimal or near optimal results for many combinatorial optimization problems~\cite{GA_ahmed2001,GA_moon2001,GA_yun2011,GA_sung2014,GA_rashid2018}. The advantage of adopting genetic algorithm is that the same solution framework is customized to solve all cases of the optimal task ordering problem, i.e., with and without precedence and conditional constraints.

The algorithm begins with a set of individuals or candidate solutions which is called a population. We define the $\mathrm{j}$-th individual as $\mathrm{\pi^j = (\pi_1^j, \pi_2^j, \dots, \pi_n^j)}$, where $\mathrm{\pi_i^j}$ is the task that executes at the $\mathrm{i}$-th position. We also define the fitness of each individual using Equation~\ref{eq:fit1} (or, Equation~\ref{eq:fit2} for conditional constraints). At each round of the algorithm, we select the best $\mathrm{K}$ pairs of individuals based on their fitness scores; and for each pair, we randomly choose a crossover point, $\mathrm{k \in \{1, 2, \cdots, n\}}$ and swap the first $\mathrm{k}$ elements of the pair to generate their offspring; and for each offspring, we perform mutation by swapping the values at two randomly chosen indices, $\{m_1, m_2\}, m_1, m_2 \in \{1, 2, \dots, n\}$; and finally, we discard all individuals that are not a valid ordering. This whole process is repeated until we reach a point when the fitness score of the best solution does not improve anymore.

\balance

\bibliographystyle{abbrv}
\bibliography{antler.bib,nwv.bib}

\end{document}